\documentclass[10pt,journal,compsoc]{IEEEtran}
\usepackage{graphicx}[pdf]
\usepackage{caption} 
\usepackage{algorithmicx}
\usepackage{algorithm}
\usepackage{algpseudocode}
\usepackage{mathtools}
\usepackage{bm}
\usepackage{esvect}
\usepackage{breqn}
\usepackage{color}
\usepackage{array}

\ifCLASSOPTIONcompsoc
  \usepackage[nocompress]{cite}
\else
  \usepackage{cite}
\fi
\ifCLASSINFOpdf
\else
\fi
\begin{document}
%
\title{Multi-Perspective Trust Management Framework for Crowdsourced IoT Services}
%
%
%
%

\author{Mohammed~Bahutair,
        Athman~Bouguettaya,~\IEEEmembership{Fellow,~IEEE,}
        and~Azadeh~Ghari~Neiat
}

%
%

\markboth{IEEE Transactions on Services Computing}%
{Bahutair \MakeLowercase{\textit{et al.}}: Multi-Perspective Trust Management Framework for Crowdsourced IoT Services}
%



\IEEEtitleabstractindextext{%
\begin{abstract}
We propose a novel generic trust management framework for crowdsourced IoT services. The framework exploits a \emph{multi-perspective trust model} that captures the inherent characteristics of crowdsourced IoT services. Each perspective is defined by a set of \emph{attributes} that contribute to the perspective's influence on trust. The attributes are fed into a machine-learning-based algorithm to generate a \emph{trust model} for crowdsourced services in IoT environments. We demonstrate the effectiveness of our approach by conducting experiments on real-world datasets.
\end{abstract}

}

\maketitle

\IEEEdisplaynontitleabstractindextext

%
\IEEEpeerreviewmaketitle

\IEEEraisesectionheading{\section{Introduction}\label{sec:introduction}}

%
%
%
%

\IEEEPARstart{T}{he} rapid development in WiFi and System-on-Chip technologies paved the way for Web-enabled devices. The interconnection between such devices led to the emergence of the Internet of Things (IoT) \cite{gubbi2013internet}. More specifically, the IoT consists of all Internet-enabled things that are connected to each other to exchange data. Such environments gave rise to a multitude of applications such as smart cities and smart homes \cite{gubbi2013internet}. An untapped application is the crowdsourcing platform whereby IoT devices crowdsource \emph{services} to other IoT devices \cite{yang2012crowdsourcing}. Services consist of two parts: \emph{functional} and \emph{non-functional} attributes. The functional attributes include the tasks that IoT devices can perform, such as processing tasks using their computing resources. The non-functional aspects of IoT services are the Quality of Service (QoS) surrounding the delivery of their functional aspects which include response time, reliability, etc. \emph{Crowdsourced services} are generally defined as services provided by the crowd to the crowd \cite{yuen2011survey}. In \emph{IoT service crowdsourcing}, IoT devices provision services to other nearby IoT devices. We refer to such services as \emph{crowdsourced IoT services}. IoT devices can crowdsource a wide range of service types such as \emph{computing resources} \cite{habak2015femto},  \emph{energy sharing} \cite{bulut2018crowdcharging}, and \emph{environmental sensing} \cite{kelly2013towards}. For example, in energy sharing services \cite{bulut2018crowdcharging}, IoT devices can wirelessly send energy to other nearby devices (service providers). IoT devices with low energy levels can consume such services to recharge themselves (service consumers).

While IoT crowdsourcing platforms provide distinct opportunities in terms of convenience and efficiency, they also present fundamental challenges. A key challenge is the ability to \emph{trust} crowdsourced IoT services. Trust is a crucial element for a successful deployment of the crowdsourcing environment. In such an environment, \emph{any} IoT device can provide services to others. A service provider could be malicious and potentially harm other IoT devices. As a result, IoT consumers would refrain from using such a platform since there are no guarantees that services would behave expectedly. The major risk at the end would be the disruption of the entire crowdsourcing ecosystem resulting in a slow (or no) adoption of the platform. In other words, the deployment of IoT crowdsourcing platforms is predicated on providing the ``right'' trust framework. For example, IoT consumers require trustworthy computing services to use computing resources from other devices. IoT computing services involve IoT devices providing their computational resources, e.g., CPU, to other nearby devices \cite{habak2015femto}. IoT consumers send their computationally-intensive tasks to service providers for processing. Consumers, however, have no guarantee that the results of their tasks are error-free, or their data has not been disclosed. These concerns can be alleviated by assessing the trustworthiness of the providers prior to service consumption. IoT crowdsourcing environments, however, exhibit specific characteristics that make trust assessment challenging. Such characteristics include the \emph{diversity} and \emph{anonymity} of IoT users and devices, and the lack of a central managing authority.

We identify two key properties in IoT platforms which necessitate trust management frameworks specifically tailored to such environments. \emph{First}, IoT is intrinsically a highly dynamic environment \cite{kyriazis2013smart}. Earlier trust frameworks relied upon previous interactions to evaluate a service's trustworthiness \cite{saied2013trust,chen2011trm}. New devices usually have no prior interactions with other devices. As a result, it is challenging to assess their services' trustworthiness. \emph{Second}, trust among IoT devices is highly dynamic and context-aware \cite{yan2014survey}. In traditional services, providers are assumed to be \emph{fixed} and \emph{known}. Earlier trust frameworks rely mainly on service providers to assess the trustworthiness of a given service. Conversely, the trust level of IoT services depends on several other elements besides service providers. IoT services are provided directly by IoT devices. Such devices exhibit unique characteristics that play a crucial role in determining the overall trustworthiness of their services. For example, IoT devices, such as \emph{wearables}, can be used by different users at any point in time (smart shoes shared between two brothers or friends). Consequently, each user may have a different impact on the overall trustworthiness of the IoT service. In that respect, the service's trust may be affected by a variety of \textit{trust indicators} including \emph{the owner's reputation}, \emph{the operating system}, \emph{the device's manufacturer}, etc. For example, an IoT device with an outdated operating system may negatively impact the trust level of its service as it may make it more susceptible to malicious attacks. In contrast, the device's owner may have a high reputation, i.e., a property that is typically linked to highly trusted services. Therefore, the \emph{multiple perspectives} of trust are paramount for establishing the trustworthiness of IoT services. In the earlier example, the trust from the \emph{device's perspective} can be low because of the outdated operating system, whereas the trust from the \emph{owner's perspective} might be high due to the high owner's reputation. It is worth noting that perspectives may not be limited to those of owners and devices. 


We develop a \emph{perspective-based} trust management framework that tries to achieve a trustworthy IoT crowdsourcing environment. The framework consists of multiple perspectives, each of which corresponds to a trust feature of IoT services (e.g., owner, device, etc.). Each perspective contributes to the trust value of a given crowdsourced IoT service. There are typically no sufficient historical records for IoT services to evaluate their trust. We address this challenge by leveraging the inherent characteristics of IoT services (e.g., providers and devices) as their trust attributes to represent their trustworthiness. Trust-related data (e.g., owner's and device's data) is exploited to assess a given service's trust. The integrity of the trust data is important but distinct from the focus of this paper.  Indeed, we assume that the IoT device’s data used in our trust model is protected by secure and privacy-preserving techniques such as blockchain, which is the subject of another parallel study.  


It is worth noting that trust is mutual. In other words, both consumers and providers trustworthiness should be ascertained to achieve a trusted platform. \emph{We focus in this work on assessing the trustworthiness of IoT service providers}.

The contribution of our work is as follows:
\begin{itemize}
    \item A \emph{perspective-based} trust management framework for assessing the trustworthiness of crowdsourced IoT services.
    \item A set of approaches and formulas for evaluating IoT services' trust from three perspectives: the owner, device, and service perspectives.
    \item A machine-learning-based algorithm for building a trust model for IoT services.
\end{itemize}

The rest of the paper is organized as follows. Section \ref{multiPerspectiveFramework} presents the proposed IoT trust management framework. Section \ref{commonPerspective} introduces the owner, device, and service perspectives. Section \ref{evaluation} discusses the experimental results. Section \ref{relatedWork} discusses related work. Section \ref{conculsion} concludes the paper.


    

\section{Perspective-based Trust Management Framework for Crowdsourced IoT Services}
\label{multiPerspectiveFramework}

We address the issue of establishing trust between IoT service providers and consumers. Service providers and consumers are assumed to own IoT devices. IoT devices are used to provide and consume IoT services. We propose a novel generic framework that assesses the trustworthiness of crowdsourced IoT services. It consists of a novel \emph{multi-perspective trust model}. Each perspective in the model (e.g., owner, devices, and service perspectives) partially influences the trust level of an IoT service. In other words, a single perspective can reflect services' trustworthiness to some extent. The aggregation of these perspectives can describe the overall trustworthiness of services. A trust perspective is ranked upon their quantitative significance. The significance indicates a trust perspective's capacity in affecting trust.

It is worth noting that the framework aims to evaluate the trust of potential IoT service providers. We use IoT service-related data (inherent characteristics of IoT services) to deduce the trustworthiness of IoT services. The correctness and integrity of such data should be ensured before assessing trust. However, in this work, we assume that the data used for trust evaluation is accurate and has not been modified.

\subsection{Motivation Scenario and Problem Definition}
We use the following motivation scenario to illustrate the significance of our work. Assume an IoT crowdsourcing environment, whereby IoT devices may offer their computing resources (e.g., CPU and memory) to other resource-poor IoT devices \cite{habak2015femto, fernando2016computing}. Suppose an IoT device owner $A$ elects to make their smartphone available to provision compute services to other nearby devices. Additionally, we assume that provider $A$ is new and therefore has not provisioned any services before. In that respect, it lacks historical data that can describe their behavior. Let us now assume that a consumer $B$ needs compute services to perform tasks on their confidential data. Provider $A$ is a potential candidate to receive consumer $B$'s task. However, consumer $B$ may not be able to use provider $A$'s services due the lack of $A$'s history, which is used to ascertain the trust consumer $B$ needs to select provider $A$. As a result, service provisioning/consumption may not occur due to the absence of trust. 

A trust management framework for IoT services should account for two challenges to successfully evaluate trust. First, IoT devices are highly dynamic. IoT devices typically have a relatively short life span, i.e., they are expected to come and go frequently. As a result, there is a high chance that an IoT service consumer would request a service from another IoT device that had just joined the IoT environment. In such a scenario, a trust management framework cannot rely on the availability of historical records of the IoT service as a result of it being fairly new. Second, trust in IoT environments is context-aware. For example, a trustworthy device does not necessitate a trustworthy service. Other \emph{context elements} may play a role in the overall service trust beside the device (e.g., the owner of the device). Therefore, the problem is defined as \emph{assessing the context-dependent trust of highly dynamic IoT services}. Our proposed framework targets both challenges by leveraging the inherent characteristics of IoT services. The framework builds a \emph{trust model} using such characteristics, which is later used to obtain the trust level of IoT services. Additionally, the context of the service is implicitly captured as a result of exploiting the IoT services' inherent characteristics.

\subsection{Multi-Perspective Trust Model}
We introduce a \emph{multi-perspective trust model} to assess the trust value of a crowdsourced IoT service $S$. Each perspective represents one of the services' properties. For example, an IoT service property could be the owner of an IoT device, which is considered as a trust perspective. A trustworthy owner increases the probability of a trustworthy service provided by their IoT device. The influence of each perspective on trust is inferred from the \emph{trust attributes} of that perspective.  For example, the social properties and locality of the owner are two attributes that reflect the owner perspective's effect on the trust. Each perspective may influence the trust value at varying degrees. For example, in some applications, an owner of a device may have a higher effect on trust than the device being used. The \emph{perspective significance} is used to reflect the level of influence a perspective has, compared to the rest of the perspectives. The generality of the proposed framework is preserved as a result of the multi-perspective trust model properties. The set of perspectives is application-based. Additionally, the significance of each perspective can have different values for different scenarios. For example, in one application, one might favor the owner perspective over the device perspective. Such a case can occur when IoT devices offer Internet access services using WiFi hotspots \cite{neiat2015spatio}. In other applications, the trust may be built by relying more on the device perspective. One example of such applications is energy sharing services \cite{bulut2018crowdcharging}. The capacity of IoT devices' batteries and their energy transmission speeds (device perspective attributes) can be more important to trust evaluation than devices' owners.

Recall that one of the main goals of the framework is to address the challenge of insufficient historical data or lack thereof. However, the framework still leverages data that \emph{indirectly} influences the trustworthiness of the IoT services (discussed in Section \ref{commonPerspective}). For instance, in the device layer (Section \ref{section:device_perspective}), the reputation of the different device properties is used to deduce a score for the device. The reputation of device properties (e.g., operating system and model) are obtained using \emph{historical records}. While such properties may not be adequate to accurately determine the trustworthiness of a service, they can be used \emph{with information from other perspectives} to increase the accuracy of the measured trust.

The trust of an IoT service $S$ is modeled as a set of perspectives $\mathcal{L}$ that influences the trustworthiness of $S$. Each perspective $L$ has a set of attributes $\mathcal{A}_L$, which governs its perspective influence on the total value of the trust. Given an attribute $A \in \mathcal{A}$, $\mathcal{S}_A$ reflects the attribute significance in influencing the trust value of the service $S$. We use the set of perspectives $\mathcal{L}$ along with their corresponding attributes $\mathcal{A}$ to devise a trust model $\mathcal{T}$ that measures the trustworthiness of any given IoT service $S$. Table \ref{tab:notations} summarizes the notations used throughout the paper.

\begin{table}[!t]
    \renewcommand{\arraystretch}{1.3}
    \caption{Summary of notations.}
    \centering
    \begin{tabular}{|c||c|}
         \hline
         Notation & Definition  \\ \hline
         $S$ & An IoT service \\ \hline
         $\mathcal{L}$ & The set of trust perspectives \\ \hline
         $L$ & A trust perspective \\ \hline
         $\mathcal{A}_L$ & The set of attributes for perspective $L$ \\ \hline
         $A$ & A trust attribute \\ \hline
         $\mathcal{S}_A$ & The significance of attribute $A$ \\ \hline
         $\mathcal{T}$ & The trust model \\ \hline
    \end{tabular}
    
    \label{tab:notations}
\end{table}

\subsection{Trust Modeling and Assessment}
Trust evaluation is achieved through a two-phase process: (1) \emph{Trust Model Preprocessing} and (2) \emph{Service Trust Assessment}. Trust model preprocessing happens \emph{offline} before service provisioning or consumption. It aims to learn the relationship between the IoT service's inherent characteristics and its trust. These relationships are inferred using previous IoT service instances. The result is a \emph{trust model} that is used in the service trust assessment phase to assess new services' trustworthiness. More specifically, the trust model is assumed to be used by consumers in a decentralized fashion to determine services' trustworthiness.

\begin{figure*}
    \centering
    \captionsetup{justification=centering}
    \includegraphics[width=0.75\textwidth]{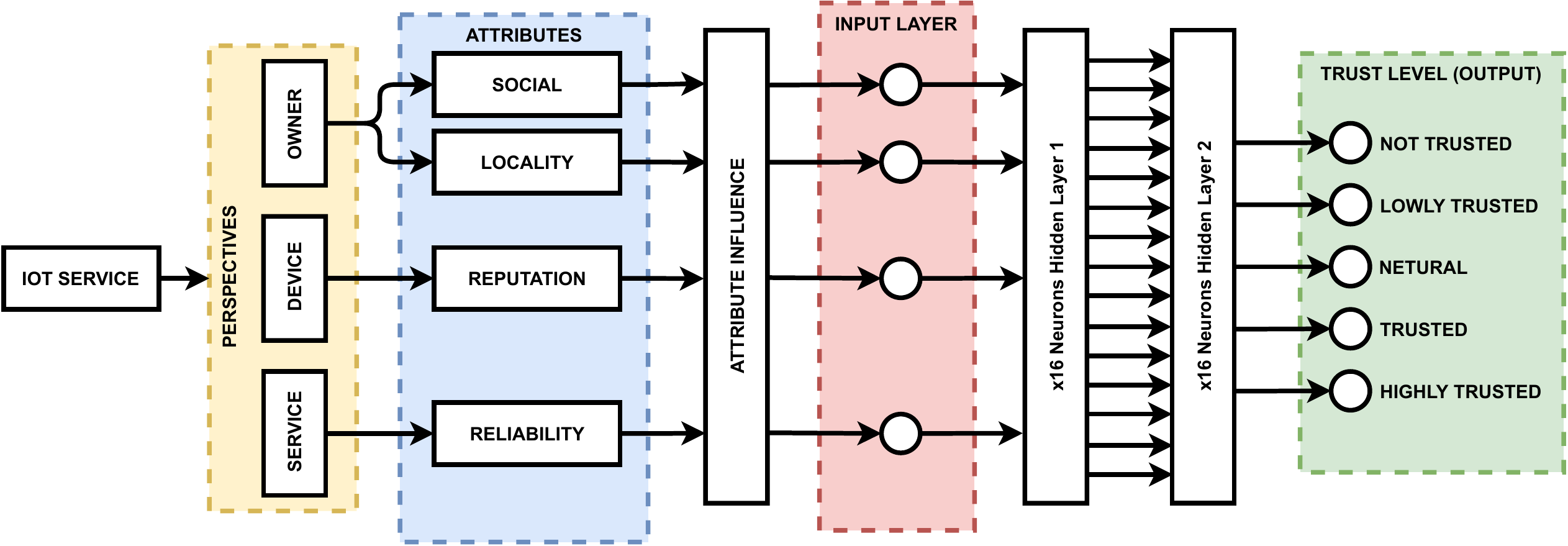}
    \caption{\centering A concrete Neural Network architecture example used for obtaining the trust model using three perspectives.}
    \label{fig:concrete_nn}
    
\end{figure*}

\subsubsection*{Trust Model Preprocessing}
The value of the trust $T$ is measured using the trust perspectives $\mathcal{L}$. The goal of the proposed framework is to identify a trust model $\mathcal{T}$ that evaluates the trust value $T$ based on the IoT service $S$ perspectives. We use Neural Networks (NN) \cite{haykin2004comprehensive}, a machine learning algorithm, to determine the model $\mathcal{T}$. Our choice of Neural Networks is based on two reasons: (1) Neural Networks can detect complex non-linear relationships between input and output variables \cite{tu1996advantages} (2) the significance of the trust attributes can easily be computed using the trained model.

One crucial aspect of the framework is its \emph{extensibility}. In other words, the framework does not target a specific service type. Different IoT service types may require additional attributes to the perspectives, or even new perspectives. To this end, the framework assesses the trustworthiness of IoT services using several perspectives. As a result, the framework \textbf{is not limited} by the perspectives and attributes discussed in Section \ref{commonPerspective}. The parameters of the framework (e.g., type and number of perspectives/attributes) is highly dependent on the type of service (e.g., WiFi hotspot, energy sharing, computing resources, etc). Changing the number of perspectives would necessitate a subsequent tweak in the neural network. Therefore, having a fixed neural network architecture would eliminate the ability to apply such tweaks.

Without loss of generality, we illustrate our model through an example of a concrete architecture using the perspectives and attributes discussed in Section \ref{commonPerspective}. Fig. \ref{fig:concrete_nn} shows how perspectives and attributes are used in our trust model. Three perspectives are used for this purpose; namely, owner, device, and service perspectives. Each perspective can have one or more attributes that influence the trust level of a service. The influence of each attribute on the trust should be computed before feeding it to the neural network. Details on influence computation is discussed in Section \ref{commonPerspective}. This example uses a 4-neuron input layer, two 16-neuron hidden layers, and a 5-neuron output layer. Each neuron in the input layer corresponds to a single attribute. The neurons in the output layer represent the trust level of a given service. In this example, a service is classified into one of five trust levels: not trusted, lowly trusted, neutral, trusted, and highly trusted.



Algorithm \ref{alg:nn} lists the steps for training the trust model $\mathcal{T}$. The algorithm takes as input the set of attributes and data samples. The algorithm also takes as an input the \emph{cost threshold} $\tau$. The cost threshold is a hyperparameter, which is set by an expert. Upon each training iteration, the trust level is computed using the trained weights. The computed trust is compared against the ground-truth (the output section of the training datasets) and the difference between them (i.e., the cost) is computed. If the cost is less than the set threshold $\tau$, the training loop is terminated. Lines (1 - 4) loop through all available attributes in each perspective and compute their contribution. These contributions differ from each other and will be discussed in Section \ref{commonPerspective} in details. The algorithm uses the evaluated attributes to train the trust model $\mathcal{T}$ (Lines 5 - 10). In our work, we use Adam Optimizer \cite{kingma2014adam} to adjust the network's weights, however, any other optimizer can be used with the algorithm. Finally, the algorithm returns the trained trust model $\mathcal{T}$ for assessing new service instances.

The trust assessment of an IoT service is carried out by substituting the attributes of each perspective in the trained neural network. However, some attributes have more influence than others, while others may have little to no influence at all on the trust. Attributes that have no influence on the trust should be filtered out and not be used for trust evaluation as it could degrade the performance of the network \cite{verikas2002feature} and the framework as a result. We achieve this by obtaining the significance of each attribute. Computing the significance $\mathcal{S}$ of the attributes is carried out using the weights of the trained NN model. We use the \emph{input-output sensitivity} formula in \cite{engelbrecht1995determining} for the significance evaluation. The input-output sensitivity formula measures the \emph{importance} of a NN's inputs over its outputs. In other words, it evaluates how sensitive NN outputs are against a given NN's input. Our aim is to evaluate the significance of the attributes (essentially, the NN's inputs) on determining the trust's level (the NN outputs). The input-output sensitivity formula is given in \cite{engelbrecht1995determining} as:
\begin{equation}
    S_{A_{L_k}} = \frac{\partial o_k}{\partial z_{A_L}}
\end{equation}
where $S_{A_{L_k}}$ is the significance of the attribute $A$ for perspective $L$ over the $k^{th}$ trust level (e.g., highly trusted). $o_k$ is the mapping function for the last NN layer, and $z_{A_L}$ is the input node corresponding to the attribute $A$ of perspective $L$. The total significance of the perspective can be computed by the following equation in  \cite{engelbrecht1995determining}: 
\begin{equation}
    S_{L} = \max\limits_{k \in K} S_{L_k}
\end{equation}
where $K$ is the total number of possible trust levels.

\begin{algorithm}
    \renewcommand{\algorithmicrequire}{\textbf{Input:}}
    \renewcommand{\algorithmicensure}{\textbf{Output:}}
    \caption{Neural Network Training for Building Trust Model $\mathcal{T}$}
    \label{alg:nn}
    \begin{algorithmic}[1]
        \Require
        $\mathcal{L}$: a set of trust perspectives,
        $N$: sample data,
        $\tau$: the threshold for the cost function
        \Ensure $\mathcal{T}$: the trained trust model

        \For{$L \in \mathcal{L}$}
            \For{$A \in \mathcal{A}_L$}
                Compute $A$'s contribution to trust
            \EndFor
        \EndFor

        \State $cost$ = $\infty$
        \State Set $A \in \mathcal{A_L}$ where $L \in \mathcal{L}$ as inputs for the Neural Network.
        \While{cost $>$ $\tau$}
            \State Optimize the Neural Network weights to the sample data $N$ using Adam's optimizer \cite{kingma2014adam}.
            \State Update the $cost$ value based on the new weights.
        \EndWhile
        \State \Return the trained model $\mathcal{T}$
    \end{algorithmic}
\end{algorithm}

\subsubsection*{Service Trust Assessment}
During service provisioning, service consumers assess services' trust using the trained trust model $\mathcal{T}$. Given a service $S$, a consumer $s_c$ computes the values of each attribute $A \in \mathcal{A}$ for every perspective $L \in \mathcal{L}$ (will be discussed in Section \ref{commonPerspective}). The computed attributes are fed into the trained model $\mathcal{T}$, which evaluates the trust $T$ for $S$.

\subsection{Confidence Assessment}
The \emph{confidence} of the trust model represents the accuracy of the measured trust. More formally, \emph{we define confidence as the probability that a framework’s result is accurate}. For example, assume our trust model assessed an IoT service as trustworthy. The \emph{confidence} of the framework indicates how \emph{certain} the assessment is. If the confidence is low, it indicates that the framework is unsure about its assessment. In contrast, a high confidence value signifies that it is highly unlikely that the service might end up being untrustworthy.

\begin{figure}
    \centering
    \includegraphics[width=0.40\textwidth]{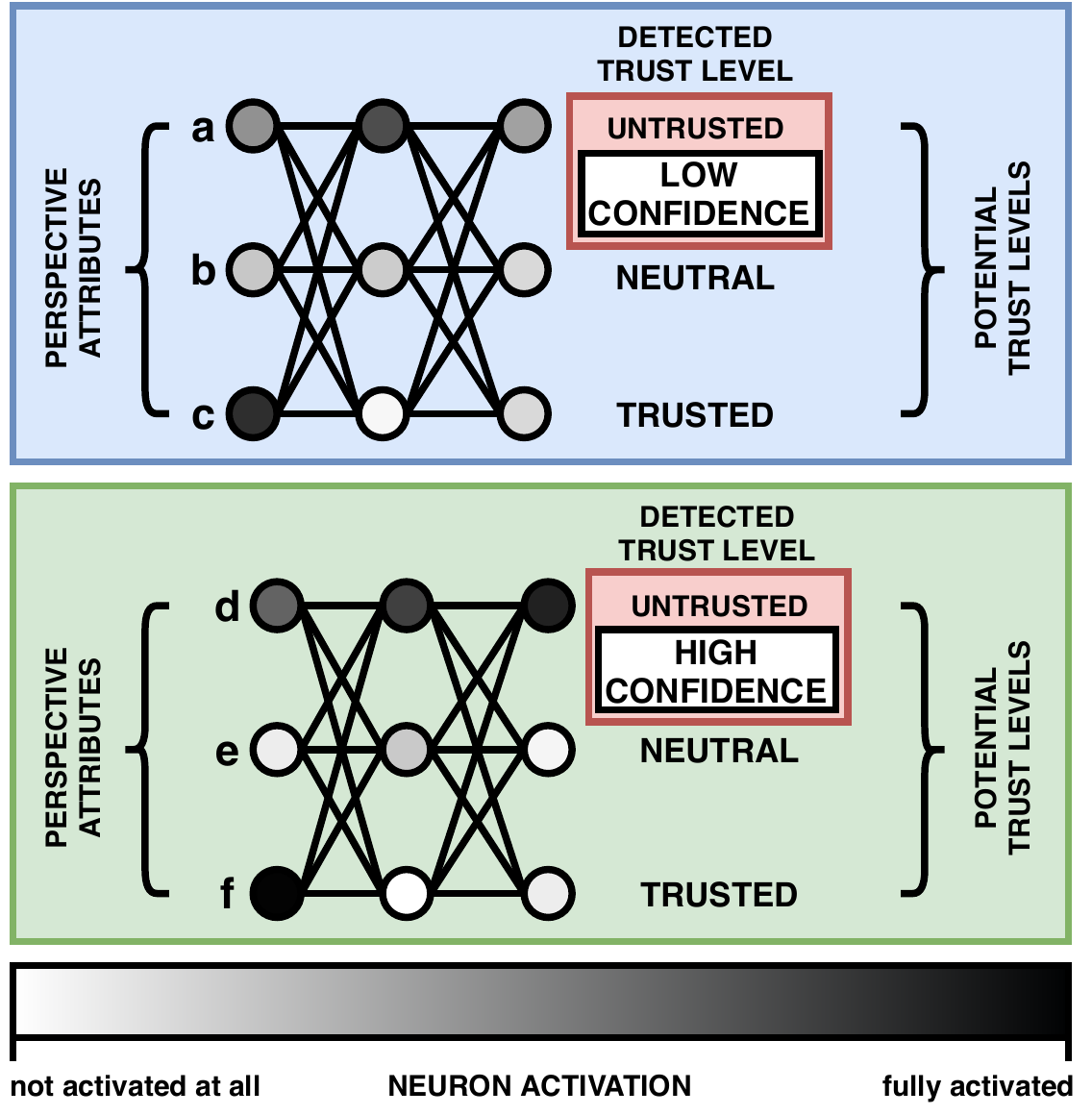}
    \caption{High confidence compared to low confidence trust assessment.}
    \label{fig:confidence_assessment}
\end{figure}

We leverage the \emph{neuron activation} in the NN output layer to obtain the confidence \cite{wan1990neural}. A Neural Network classifies input samples based on neurons activation in the output layer \cite{haykin2004comprehensive}. The neuron activation is obtained after substituting the results of previous layers into its mapping functions. Given an input sample to classify, the output neuron with the highest activation value is selected. The input sample is then classified into the class associated with the selected output neuron. Fig. \ref{fig:confidence_assessment} illustrates an example of a trained Neural Network used in two different scenarios. The circles represent the neurons, and the level of shade indicates its activation. A blacked out circle corresponds to a fully activated neuron. For the sake of simplicity, the network takes as input three perspective attributes to classify a service's trust level to either untrusted, neutral, or trusted. We consider that the correct assessment for the service is ``untrusted'' for this example. In both scenarios, the network assesses the service as ``untrusted''. Note that although both cases produce the same result, the activation value is considerably different. The first set of inputs produce only a half activated ``untrusted'' output neuron. Additionally, the other two output neurons are noticeably activated making it difficult to distinguish between them and the ``untrusted'' output neuron. The second set of inputs produces a fully activated ``untrusted'' output neuron. The other output neurons are almost deactivated. We consider the assessed trust in the first scenario as \emph{untrusted with low confidence} while the second scenario as \emph{untrusted with high confidence}.

We use the \emph{softmax function} \cite{nasrabadi2007pattern} to extract the confidence information from the network. As mentioned earlier, a neural network determines the class of a given input based on the activation of the output layer’s neurons. In other words, the neuron with the highest activation is chosen as the predicted class for any given input. The activation value of a neuron is computed using an activation function. There are several activation functions that can be used, e.g., sigmod, and softmax. The activation value can vary by changing which function is used. One distinctive property of the softmax function is that the computed activation values are in fact \emph{probabilities}. In other words, each activation value answers the following question: “what is the probability that a class is the right choice given an input?”. Since they are probabilities, the activation values of the output neurons always sum to one. For example, if a neural network has three outputs representing classes A, B, and C, then a softmax function might compute the activation values 0.8, 0.15, and 0.05 for them, respectively. Other activation functions (e.g., sigmod) may calculate the following activation values 0.7, 0.35, 0.25 for the three classes, respectively. By using the softmax function, we can leverage the value of the selected neuron as a measure to determine how \emph{confident} the network was when it predicted the class associated with that neuron. Given the $i^{th}$ neuron's activation value $x_i$, the softmax function can be expressed as:

\begin{equation}
    f(x_i) = \frac{e^{x_i}}{\sum\limits_{i=1,2,3...}e^{x_i}}
\end{equation}

Computing the trust confidence is achieved by substituting the activation value of the detected neuron in the softmax equation.

\section{Trust Perspectives}
\label{commonPerspective}
The proposed framework evaluates the trust based on the inherent characteristics of IoT services. Recall that an IoT service is transformed into a multi-perspective trust model using their inherent characteristics. Each perspective contributes partially to the service's trust. Perspectives are application-dependent and can vary across different scenarios. In this section, we identify three perspectives that may be used in most application scenarios: (1) \emph{Owner Perspective}, (2) \emph{Device Perspective}, and (3) \emph{Service Perspective}. Our framework leverages IoT services' data to assess their trust level. The privacy and integrity of the data are crucial but is not the focus of this work. We, therefore, assume that the data is protected using integrity and privacy-preserving techniques such as blockchain, which can be further studied in future work.

\subsection{Owner Perspective}
The owner perspective contributes to an IoT service's trust based on the service provider's (device owner) trustworthiness. The level of influence of this perspective depends on its attributes. We propose two attributes: \emph{social relations} and \emph{locality attributes}.

\subsubsection*{Social Relations Attribute}
The social relation between an IoT service provider $s_p$ and consumer $s_c$ can affect the level of trust between them \cite{gilbert2009predicting}. Such relations are leveraged to infer the level of trust between users. We assume social relations are accessible through public social profiles (e.g., LinkedIn and Facebook) or shared private information. We first start with public profiles. Private information can be trusted with the system to optimize the calculation of the trust in case of unavailable public information. Users in social networks often engage in relationships with others. Some social networks differentiate between relationships based on their strength, e.g., Facebook has family, friend, or coworker relationships.

The trustworthiness of a service can be deduced using a provider's and consumer's social relation. For instance, assume a crowdsourcing environment where IoT devices share their Internet using WiFi hotspots. Suppose a consumer who wishes to use one of the available WiFi hotspots services. The consumer might trust service providers that they know (e.g., friends on Facebook) rather than a complete stranger. As stated earlier, trust attributes can vary and their significance can change depending on the IoT service type. For example, social relations could be used in WiFi hotspot services, but its significance may not be as high for environmental sensing or energy sharing services. We discuss more appropriate attributes in later sections.

Several factors affect the trust in terms of social properties two of which are: the \emph{provider's and consumer's relation}, and \emph{common friends between them} \cite{gilbert2009predicting}. We use the following equation to evaluate the \emph{relationship factor} $R_{s_p, s_c}$:

\begin{equation}
    R_{s_p, s_c}=\frac{strn_{s_p, s_c}}{K}
\label{r}
\end{equation}

where $strn_{s_p, s_c}$ is the strength of the relationship between $s_p$ and $s_c$. The strength in our work is extracted from the relation type provided by social networks (e.g., family, friend, colleague, etc). The relationship strength can be a value from 1 (weak relation, e.g., colleague) to $K$ (strong relation, e.g., family), where $K$ is the number of available relationship types (e.g., $K = 3$ if available relation types are family, friend, and colleague).

As stated earlier, the trust between a provider and consumer can also be affected by common friends between them. Let the sets $F_{s_p}$ and $F_{s_c}$ represent the friends for $s_p$ and $s_c$, respectively. Given  $F = F_{s_p} \cup F_{s_c}$, we use Eq. \ref{cf} to measure the \emph{common friend's factor} $CF_{s_p, s_c}$ between $s_p$ and $s_c$:

\begin{dmath}
    CF_{s_p, s_c}= \frac{1}{|F|} * \sum\limits_{f \in F_{c, p}} \mu_1\frac{strn_{s_p, f}}{K} + \mu_2\frac{strn_{s_c, f}}{K}
\label{cf}
\end{dmath}
where $\mu_k$ is a weighting factor and $\sum\limits_{k = 1}^{2} \mu_k = 1$

\subsubsection*{Locality Attribute}
Friends on social networks can be divided into two groups: those who had face-to-face contact, and those who know each other only on the network. For instance, assume Adam has two friends on Facebook; Ben and John. Adam knows Ben only on Facebook. John, however, lives in Adam's neighborhood. There is a high chance that Adam and John have met and spent time in person due to their close proximity. Therefore, their relationship may be stronger. \emph{The intuition is friends who met each other in real-life tend to have a stronger relation than those who have not met before} \cite{pelechrinis2012location, wang2011human}. We refer to such a concept as \emph{face-to-face factor}. 

We define the \emph{locality} $loc_u$ of a user $u$, which represents the spatial properties of the user. The locality can be a hierarchical set of attributes or a simple attribute. The user's locality provides a description of the area they belong to. For example, one way to represent the user's locality is \emph{\{Empire State Building, New York, USA\}}. Longitude and latitudes can also be used to represent the user's locality. Given the localities for $s_p$ and $s_c$ as $loc_{s_p}$ and $loc_{s_c}$, respectively, the face-to-face factor is incorporated into Eq. \ref{r} to optimize its accuracy:

\begin{equation}
    R_{s_p, s_c}=\frac{strn_{s_p, s_c}}{K} * FF(loc_{s_p}, loc_{s_c})\label{r_ff}
\end{equation}
where $FF(loc_{s_p}, loc_{s_c})$ is the probability two users have met given their localities. The way $FF$ is evaluated can vary. The $FF$ in this work is calculated based on the coordinates of each locality. Coordinates with shorter distances give a high value for the $FF$. In other words, $FF$ is measured as follows:

\begin{dmath}
    FF(loc_u, loc_v)= 1 - \frac{1}{2}\left(\frac{|loc_{u_x}-loc_{v_x}|}{max(loc_{u_x},loc_{v_x})} + \frac{|loc_{u_y}-loc_{v_y}|}{max(loc_{u_y},loc_{v_y})}\right)
\label{ff}
\end{dmath}
where $loc_{u_x}$ and $loc_{u_y}$ are the coordinates of the locality. The value of $FF$ is bound between 1 and 0. Higher $FF$ values indicate a higher chance of face-to-face contact. Similarly, the localities can be incorporated into Eq. \ref{cf}:

\begin{dmath}
    CF_{s_p, s_c}= \frac{1}{|F|} * \sum\limits_{f \in F_{c, p}} \mu_1\frac{strn_{s_p, f}}{K}*FF(loc_{s_p}, loc_f) + \mu_2\frac{strn_{s_c, f}}{K}*FF(loc_{s_c}, loc_f)
\label{cf_ff}
\end{dmath}

\subsection{Device Perspective}
\label{section:device_perspective}
The device perspective encompasses all properties related to the IoT device itself (e.g., device manufacturer and operating system). The contribution of the perspective reflects the influence of the device's properties on the IoT service's trustworthiness. As stated earlier, attributes may change depending on the service type. Additionally, the significance of the attributes can vary across different service types. Device perspective attributes can have more impact on the trust in services where the device and its performance is crucial (e.g., sensing and energy sharing services). For instance, assume a crowdsourcing environment where IoT providers share their energy with other IoT devices. A device capability and efficiency in harvesting, preserving, and sending energy (device properties) could have a higher influence on trust than owner-related properties. Device perspective attributes might have less effect on trust for services where device properties are less crucial (e.g., WiFi hotspot services). We propose the \emph{device reputation} attribute  for the device perspective.

\subsubsection*{Device Reputation Attribute}
The device reputation can affect the overall trustworthiness of the IoT service. For example, a reputable manufacturer could lead to a higher probability of reliability and less susceptibility to attacks. On one hand, a reputable manufacturer often updates their devices' firmware for increased performance and security. On the other hand, a disreputable manufacturer might push delayed updates or none at all which puts their devices at risk. The purpose of the perspective is to extract the influence of a device properties on trust.

An IoT device $d$ is described as a property set $P$, where $p \in P$ corresponds to a property of the device $d$. A property of an IoT device can be its manufacturer, operating system, type (e.g., smartwatch or smartphone, etc.), etc. The reputation $\mathcal{R}$ of a device $d$ is the reputation aggregation of all properties in the set $P$. In other words, the reputation of an IoT device is evaluated based on the reputation of the device's properties. For example, an IoT device is considered to have a high reputation if its manufacturer and operating system are also highly reputable. Similarly the type of the device can affect a service's trust level. For instance, in energy sharing services, smartshoes that harvest energy while walking might be more reliable to share energy than a smartwatch with small batteries. The reputation $\mathcal{R}$ of a device $d$ can be computed as follows:

\begin{equation}
    \label{eq:device_reputation}
    \mathcal{R} = \frac{1}{|P|}\sum\limits_{p \in P}R_p
\end{equation}
where $R_p$ is the property $p$'s reputation, and $|P|$ is the number of IoT device's properties. In the equation, the overall reputation of the device is computed based on the reputations of its properties. The average of the properties' reputation is taken as the reputation of the IoT device. It is worth noting that in this work, it is assumed that the reputation data for a device's property is available. However, if such data is missing, the problem becomes a \emph{bootstrapping} problem. Other bootstrapping techniques should be employed in such cases, which is outside the scope of this paper.

In Eq. \ref{eq:device_reputation}, the average of the properties' reputation is taken to obtain the reputation of the device. Treating all properties equally may not give accurate results. For example, assume a device $A$ which has a reputable manufacturer and runs an outdated operating system. Additionally, assume that the operating system has been used by a wide range of devices, whereas the manufacturer has produced only a few devices (i.e., less popular). The reputation of the operating system should weigh more than the manufacturer's since it has been used and reviewed by a larger number of users. Therefore, the reputation of the IoT device $\mathcal{R}_d$ is reformulated as follows:

\begin{equation}
    \label{eq:device_reputation_with_popularity}
    \mathcal{R}_d = \frac{1}{\sum\limits_{p\in P} Pop_p}\sum\limits_{p \in P}R_p*Pop_p
\end{equation}
where $Pop_p$ is the \emph{popularity} of the property $p$ which refers to the number of times a device with $p$ has been used.

\subsection{Service Perspective}
The service perspective captures the properties of a service during its operation (e.g., reliability). For example, in a crowdsourcing where users share their computing resources, the service perspective tries to measure a service's performance and its effect on trust. A service that provides fast processing for tasks will increase its trustworthiness because of the increased reliability. Otherwise, the service's trustworthiness should be lowered accordingly (i.e., a service consumer may trust the service less to process its tasks quickly due to the decreased reliability). We propose the \emph{reliability attribute} in the service perspective.

\begin{figure*}
    \centering
    \captionsetup{justification=centering}
    \includegraphics[width=0.85\textwidth]{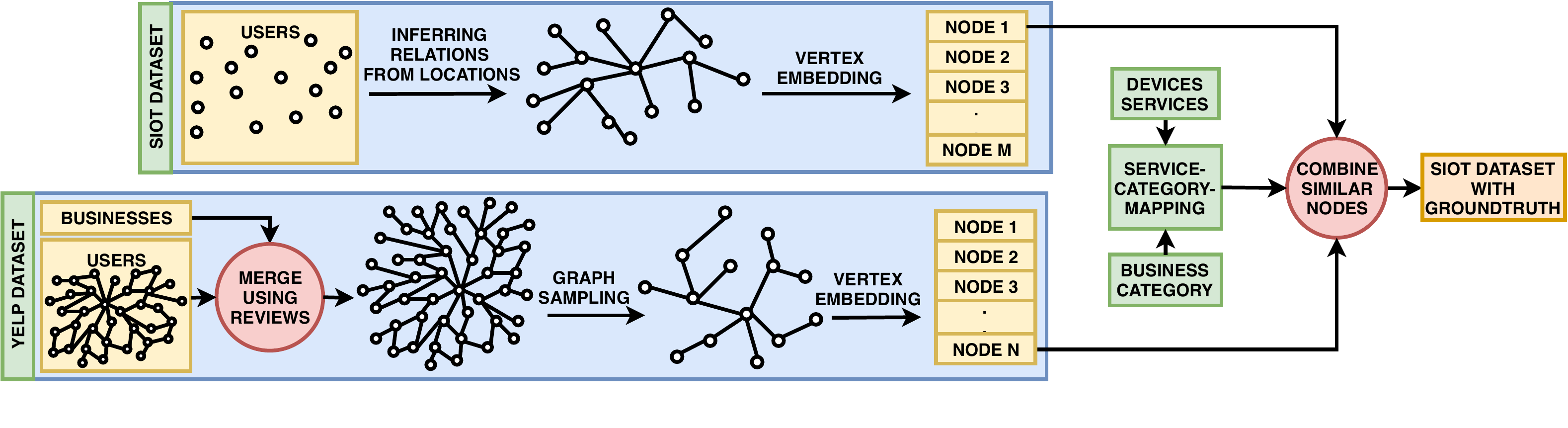}
    \caption{\centering Augmenting Yelp review rates on SIoT dataset}
    \label{fig:dataset_merge}
    
\end{figure*}

\begin{figure*}
    \centering
    \begin{minipage}{.37\textwidth}
        \centering
        \vspace{0.5cm}
        \includegraphics[width=\textwidth]{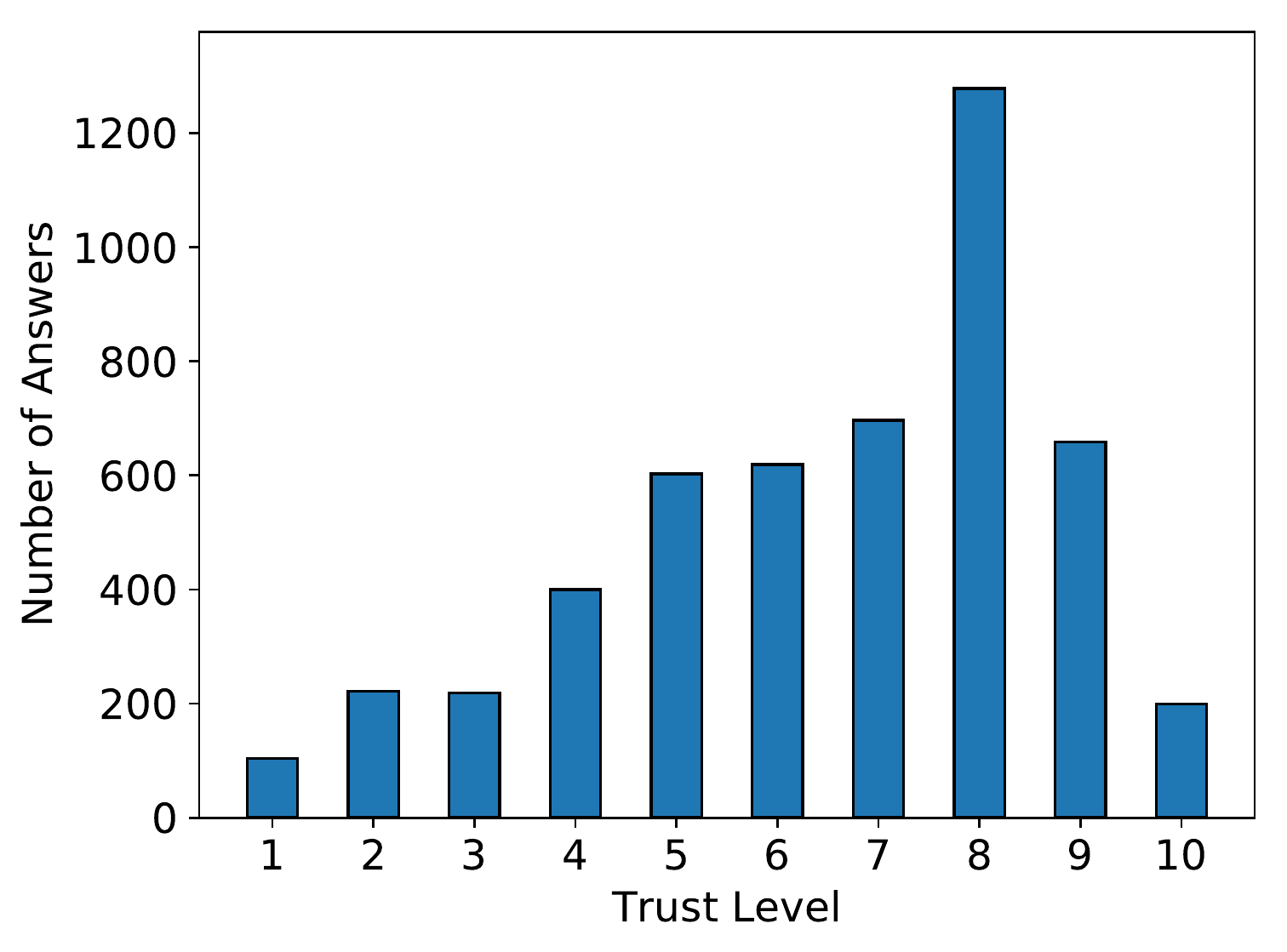}
        \vspace{-0.5cm}
        \caption{The distribution of mTurk workers' answers}
        \label{fig:answers_distribution}
    \end{minipage}
    \hspace{1.5cm}
    \begin{minipage}{.40\textwidth}
        \centering
        \includegraphics[width=\textwidth]{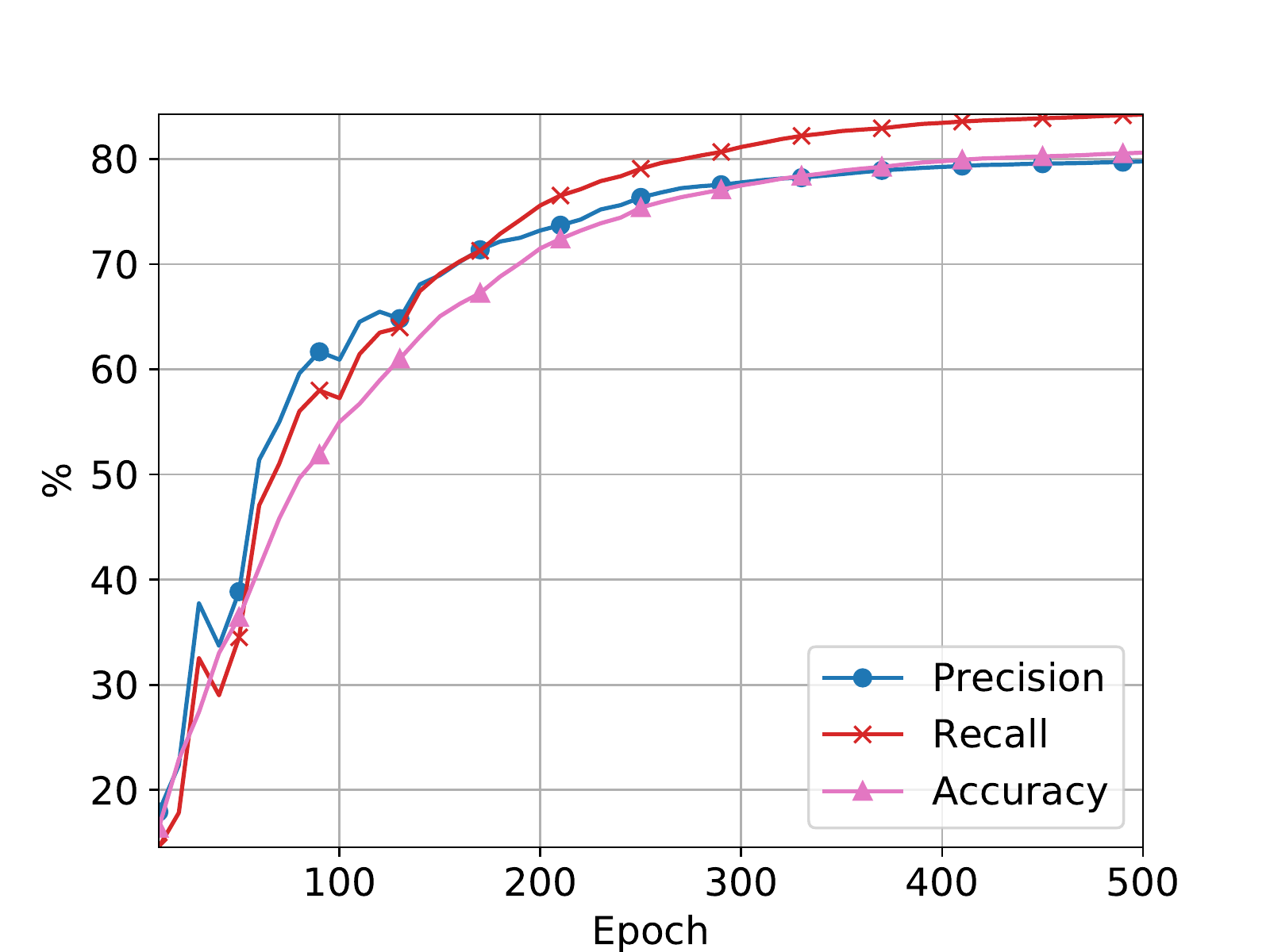}
        \caption{Precision, recall, and accuracy for the original dataset.}
        \label{fig:pra_original_dataset}
    \end{minipage}
\end{figure*}

\subsubsection*{Reliability Attribute}
We use the reliability attribute to capture the current and dynamic reliability of a service $S$. Before service provisioning, we assume that a service provider compiles a \emph{claim vector} $\vv{\bm{C_{S}}}$ which represents the performance level they claim to provide. For example, a compute service provider may list their claims as being in the area for 3 hours, and having 3 processing cores and 4GB of RAM for sharing. When a service provider decides to provision a service, we assume that they broadcast their services to nearby potential consumers. During service broadcasting, the service provider also sends their claim vector $\vv{\bm{C_{S}}}$.

Throughout service consumption, each consumer $s_c$ continuously compares the broadcasted claim vector $\vv{\bm{C_{S_{brod}}}}$ with the actual performance they get  $\vv{\bm{C_{S_{actual}}^t}}$ at time $t$. Consumer $s_c$ computes a \emph{reliability vector} $\vv{\bm{RV_{c}^{t}}}$ at time $t$ using Eq. \ref{eq:reliability_vector}

\begin{equation}
    \vv{\bm{RV_{c}^{t}}} = 1 - \left[(\vv{\bm{C_{S_{brod}}}} - \vv{\bm{C_{S_{actual}}^t}})\oslash(\vv{\bm{C_{S_{brod}}}})\right]
    \label{eq:reliability_vector}
\end{equation}
where $\oslash$ is the pairwise division operation. 

Each element in $\vv{\bm{RV_{c}}}$ corresponds to a certain claim. Elements in  $\vv{\bm{RV_{c}}}$ take values between 0 and 1 to represent how well their corresponding claims are fulfilled for consumer $c$ (i.e., 0 not fulfilled at all and 1 completely fulfilled).

Service $S$ consumers update an \emph{accumulated reliability vector} $\vv{\bm{RV_{{acc}_c}}}$ that summarizes fulfillment of the service provider claims. $\vv{\bm{RV_{{acc}_c}}}$ is updated upon the computation of a new reliability vector $\vv{\bm{RV_{c}^{t}}}$ at time $t$ using Eq. \ref{eq:acc_reliability_vector}.

\begin{equation}
    \vv{\bm{RV_{{acc}_c}}} = \gamma \vv{\bm{RV_{{acc}_c}}} + (1-\gamma) \vv{\bm{RV_{c}{t}}}
    \label{eq:acc_reliability_vector}
\end{equation}
where $\gamma$ is weighting factor between zero and one. A high $\gamma$ value results in favoring the accumulated reliability over current observed reliability.

Before a new consumer starts service consumption, it requests available accumulated reliability vectors. The consumer computes the \emph{overall reliability vector} $\vv{\bm{RV_{all}}}$ of $S$ by aggregating the set of reliability vectors. Each vector is weighted based on the time its corresponding consumer has spent consuming the service (Eq. \ref{eq:all_reliability_vector}).

\begin{equation}
    \vv{\bm{RV_{all}}} = \frac{\sum\limits_{c \in C} t_{spent_c} *  \vv{\bm{RV_{{acc}_c}}}}{t_{spent_{max}}}
    \label{eq:all_reliability_vector}
\end{equation}
where $t_{spent_c}$ is the consumption duration for consumer $c$ and $t_{spent_{max}} = \max\limits_{c \in C} t_{spent_c}$.

Finally, the service's reliability is determined as follows:

\begin{equation}
    R_S = \frac{\sum\limits_{r \in \vv{\bm{RV_{all}}}} r}{\left|\vv{\bm{RV_{all}}}\right|}
    \label{eq:service_reliability}
\end{equation}

We illustrate the reliability attribute using the following example. Assume a user $A$ wishes to use their smartphone to provide compute services. User $A$ claims that their service has 4 cores and 4GB of RAM, and would provision their service for two hours. User $A$ starts broadcasting their service and his list of claims. Assume another user $B$ in the proximity of user $A$ that wants to consume a compute service using their smartwatch. User $B$ sees a broadcasted service from user $A$ and its claims list. Since user $B$ is the first consumer, the service reliability has a \emph{default value}, say 0.5.  Once user $B$ begins consuming the service, it starts monitoring and comparing the actual and claimed reliabilities. The results of the comparisons are stored and accumulated over time. Assume another consumer $C$ that wishes to consume the service provided by $A$. Potential consumer $C$ asks current consumers for their accumulated observed reliability. Consumer $C$ uses $B$'s observed reliability as the basis to evaluate the service's reliability. When user $C$ starts the consumption, the service would have two observers that can later contribute to evaluate its reliability for new consumers.

\section{Evaluation}
\label{evaluation}
We conduct four sets of experiments to evaluate the proposed approach in terms of accuracy and training time. The first set examines the accuracy in detecting the level of trust between providers and consumers. The second set investigates the accuracy of each individual trust perspective in detecting trustworthy IoT services. The third set studies the trust model training time. The final set compares our approach to the approach proposed in \cite{bao2012dynamic}. 

\begin{figure*}
    \centering
    \begin{minipage}{.40\textwidth}
        \centering
        \includegraphics[width=\textwidth]{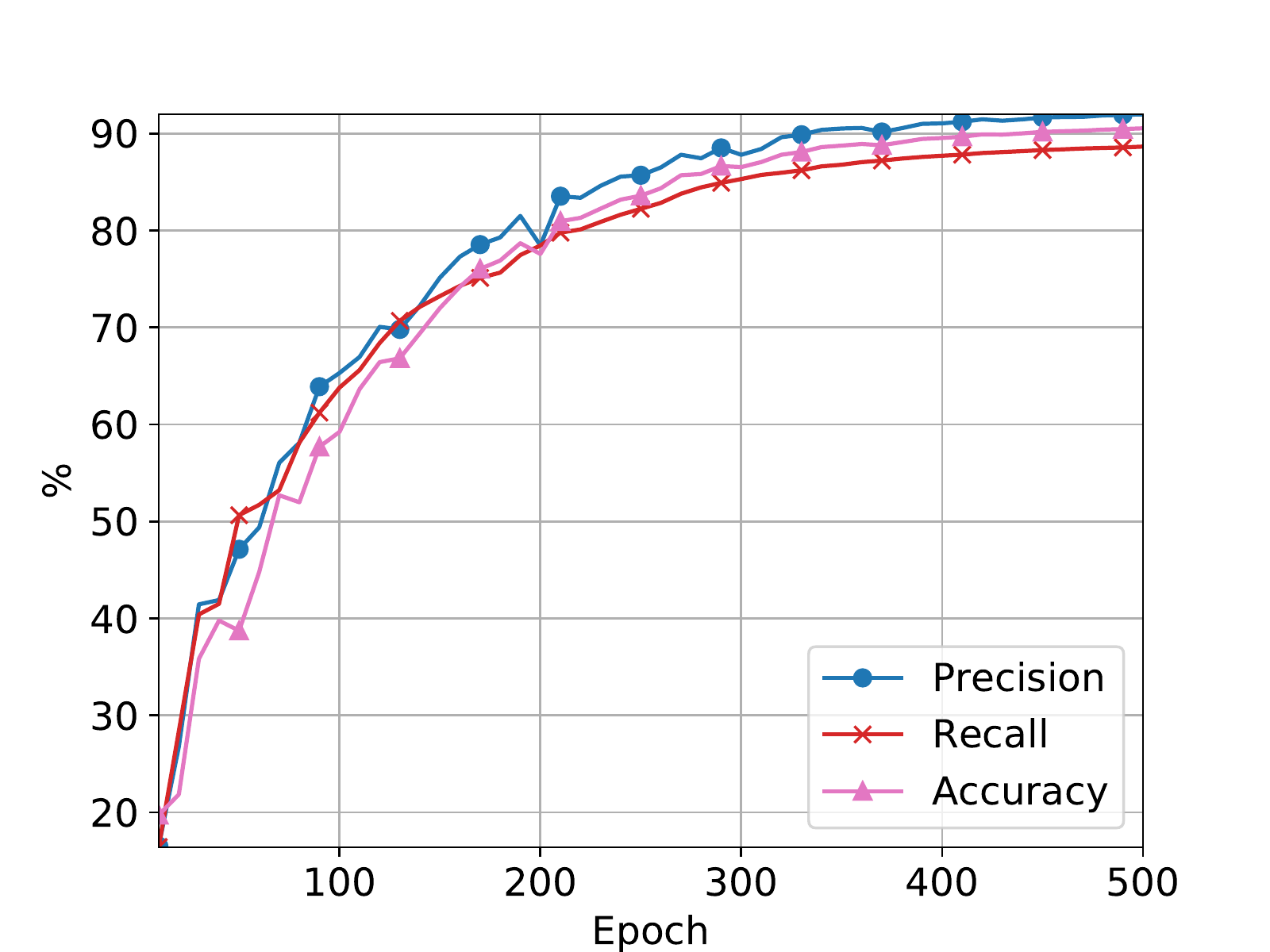}
        \caption{Precision, recall, and accuracy for the interpolated dataset.}
        \label{fig:pra_interpolated_dataset}
    \end{minipage}
    \hspace{1.5cm}
    \begin{minipage}{.40\textwidth}
        \centering
        \includegraphics[width=\textwidth]{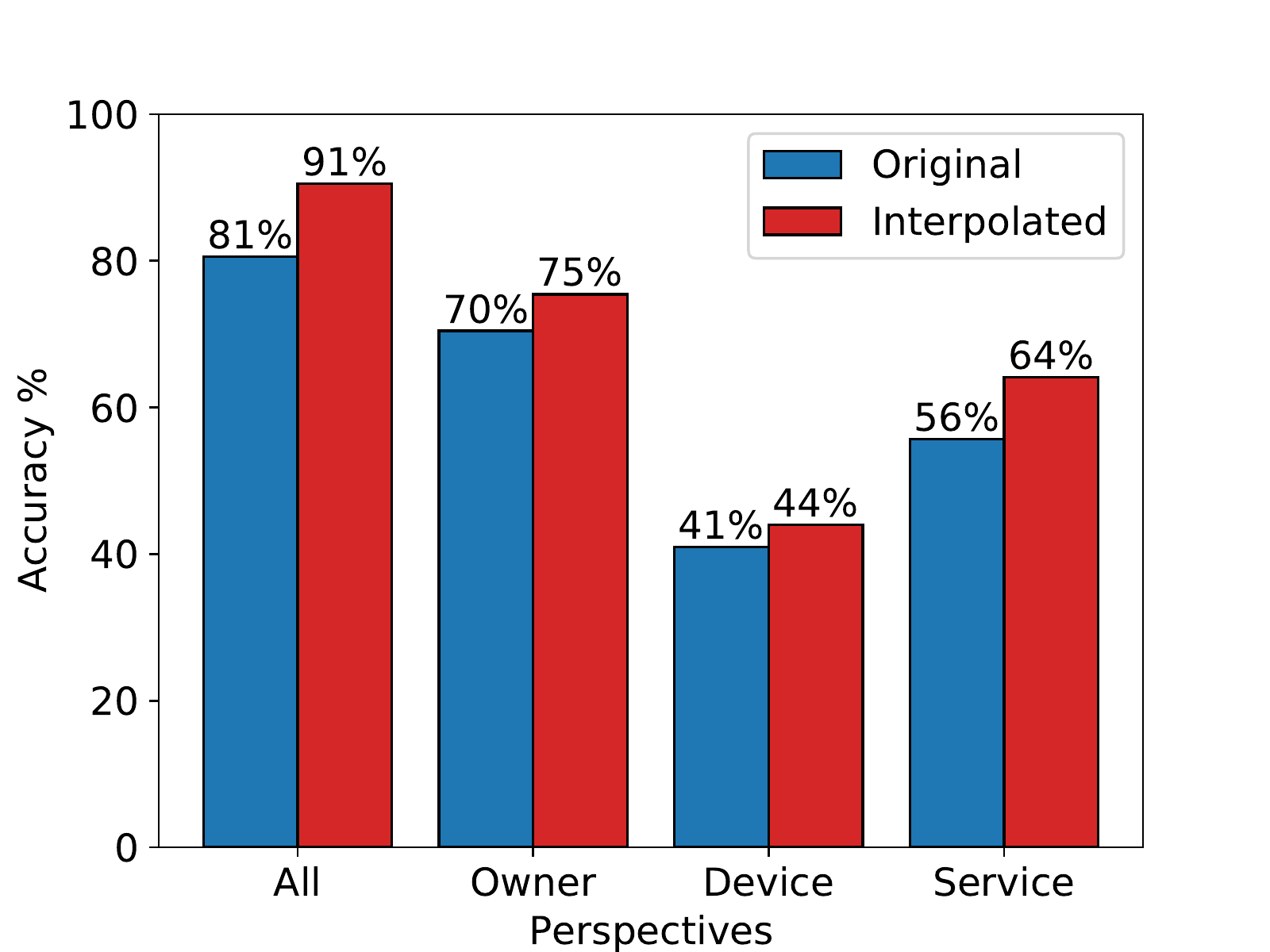}
        \caption{Per-perspective accuracy for original and interpolated datasets.}
        \label{fig:per_perspective_accuracy}
    \end{minipage}
\end{figure*}

\subsection{Dataset Description}
\label{section:dataset_description}
We use the crowdsourcing platform Amazon Mechanical Turk\footnote{https://www.mturk.com/} (MTruk) to collect the dataset for our experiments\footnote{This dataset was collected due to the absence of any publicly available IoT crowdsourcing environment dataset.}. Several questionnaires are designed to be answered by Mechanical Turk workers. Each questionnaire starts by presenting the IoT crowdsourcing environment to workers. We adopt a crowdsourced WiFi hotspot environment where, given a specific area, users can share/use WiFi hotspots to/from other IoT devices. MTurk workers are asked to consider themselves as service consumers. Additionally, MTurk workers are asked to assume that a potential WiFi hotspot service is available. A potential service is portrayed as \emph{a list of attributes} that are later grouped into \emph{trust perspectives}. The attributes used in the questionnaires include: social relation, owner reputation, device brand, device model, device operating system, concurrent consumers, and carrier reputation. Workers are asked to use their best judgment to assess the trustworthiness of a given potential service based on its attributes. Workers submit their assessment by giving a value between 1 and 10, 1 indicates an untrustworthy service, whereas 10 represents a highly trusted service. A total of 5000 questionnaires are created and published on MTurk.  Each questionnaire presents a new potential service by varying the service attributes. The distribution of the workers answers is shown in Fig. \ref{fig:answers_distribution}. Additionally, we use the collected dataset to generate a larger dataset using \emph{linear interpolation}. We use linear interpolation as it preserves the consistency of the dataset \cite{pownuk2017linear}. The new interpolated dataset is generated to test the framework's scalability for larger datasets. Moreover, the proposed framework leverages a machine learning-based algorithm that requires a larger dataset for better results. The dataset is expanded to 50,000 samples. We show the results from both datasets; i.e., original and interpolated datasets.

We analyze the answers provided by the workers to ensure honest answers and detect outliers. Specifically, we ascertain accurate answers by (1) analyzing the time it took a worker to answer a survey, and (2) asking multiple workers to answer a single unique survey. Our surveys take roughly one to two minutes to read, understand, and answer. Therefore, a worker's answer is dismissed if they take a period outside that range. More specifically, we dismiss all workers that answer the survey in less than one minute and more than five minutes. Additionally, every unique survey is presented to 10 different workers. The answers from the 10 workers are examined to ensure they are somewhat similar. The answers are dismissed if the difference between the mean of answers and any of the 10 answers is more than two units. In case of a dismissal, the survey is presented to 10 new workers. In case of approval, one answer is taken randomly as the answer to the survey.

\subsubsection*{Yelp and SIoT Datasets:} In our experiments, we compare our proposed framework with the approach proposed in \cite{bao2012dynamic}. The approach in \cite{bao2012dynamic} measures the trustworthiness of a service based on its historical records. Our collected MTurk-based dataset does not include previous trust values for a specific service. As a result, it is not suitable to compare the two approaches using the collected dataset. We, therefore, use a dataset that includes such information while having some similarity to IoT crowdsourcing environments. Two real datasets have been used for this scenario: (1) a Social IoT (SIoT) dataset \cite{marche2018dataset}, and (2) Yelp dataset\footnote{https://www.yelp.com/dataset}. The SIoT dataset consists of 16,216 IoT devices of which 14,600 belong to private users and 1,616 from public services. Device features (e.g., type and brand) are also included in the dataset. The total number of device owners is 4,000. Additionally, the dataset lists the available services provided by each device such as traffic status. Yelp is a social network whereby users rate and review businesses (e.g., restaurants). Users can also have friendships with other users. The dataset contains approximately 1,600,000 users, 6,600,000 reviews, and 192,000 businesses.

We use the SIoT dataset as our main dataset and augment it with the rates from Yelp's dataset. The rates will act as trust levels of the services provided by IoT devices in the SIoT dataset. We use graph theory to merge the two sets, see Fig. \ref{fig:dataset_merge}. The steps we take to achieve the merge are as follows: (1) Convert the SIoT dataset into a graph, (2) Incorporate businesses into Yelp's users' graph, (3) Perform graph sampling on Yelp's dataset, and (4) Perform the merge between the SIoT and Yelp graphs.

The SIoT dataset includes location information about devices and their owners. We use these locations to infer relationships between users \cite{pham2016inferring}. The result is a graph of users connected with weightless edges. Relationships between users exist in Yelp's dataset. We construct a new graph where it includes users and \emph{businesses} as the nodes of the graph. We use users' reviews to build links between users and businesses. Review rates are used as weights between users and businesses. The resulting graph consists of nodes with a set of weighted and weightless edges. The remaining challenge is the large difference in size between the two graphs: the SIoT graph has 4,000 nodes and Yelp's graph has around 1.5 million nodes. We perform graph sampling \cite{leskovec2006sampling} on Yelp's graph to reduce its size. We then weight edges in the SIoT graph by finding matching edges from the sampled Yelp and SIoT graphs using graph embedding \cite{yan2007graph}. Generally, graph embedding is a technique to transform graph vertices into vectors. Moreover, we leverage businesses' categories provided by Yelp's dataset to make the matching more accurate. We map the set of categories to the set of available services in the SIoT dataset. The edges' weights in the resulting SIoT graph are considered as the ground truth.

\begin{figure*}
    \centering
    \begin{minipage}{.40\textwidth}
        \centering
        \includegraphics[width=\textwidth]{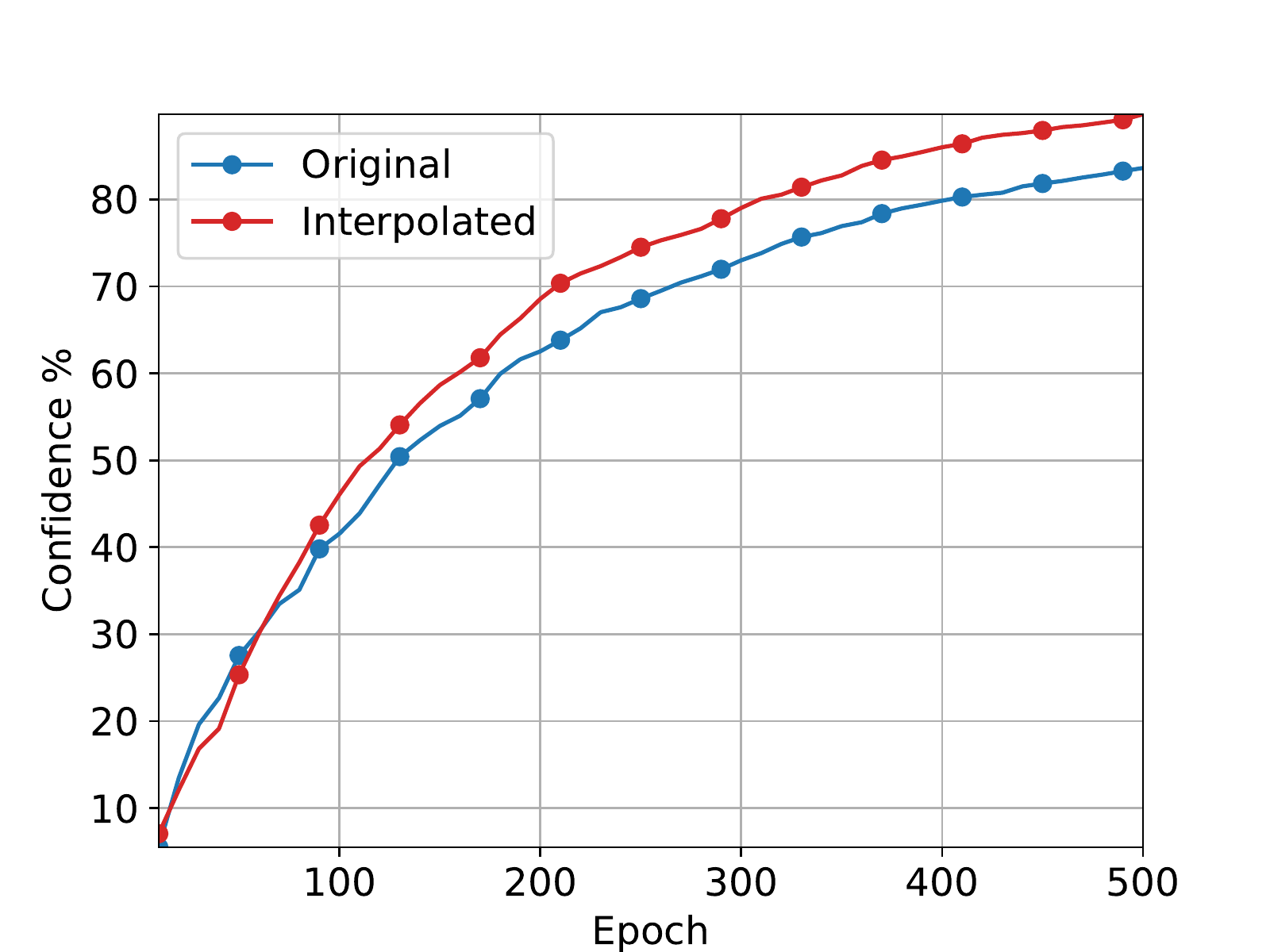}
        \caption{Framework's confidence for original and interpolated datasets.}
        \label{fig:confidence}
    \end{minipage}
    \hspace{1.5cm}
    \begin{minipage}{.40\textwidth}
        \centering
        \includegraphics[width=\textwidth]{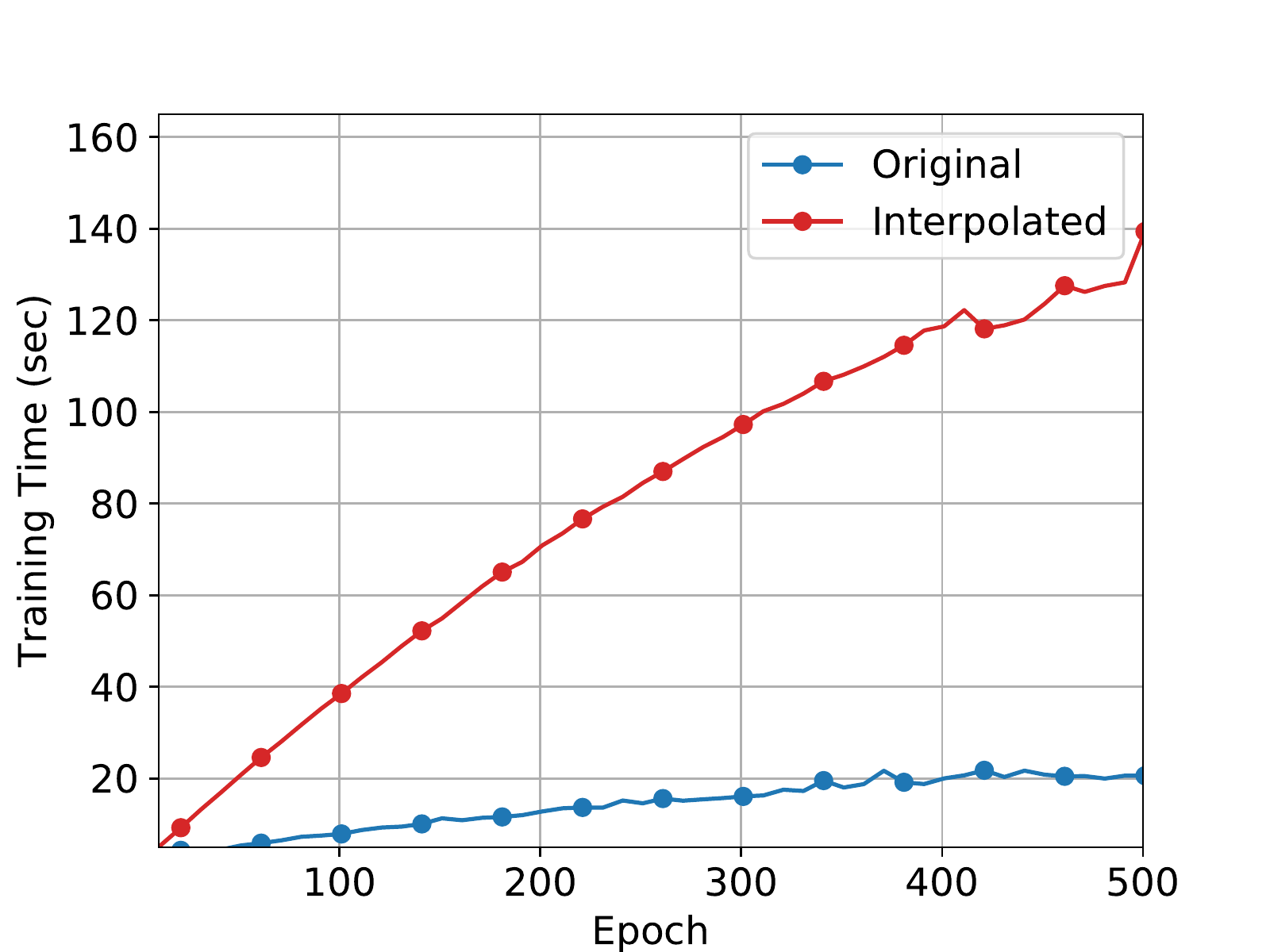}
        \caption{Trust model training time for original and interpolated datasets.}
        \label{fig:training_time}
    \end{minipage}
\end{figure*}

\subsection{Experiment Setup}
The aim of our experiments is to examine the accuracy of the proposed framework. More specifically, we assess the framework's accuracy in detecting the ``right'' trust level of a given service. We use the answers given by MTurk workers as our baseline. In other words, the accuracy of the framework is obtained by comparing the computed results with the answers of the MTurk workers.

We use \textasciitilde{5000} instances of service provisionings to evaluate the accuracy of our framework. The data is split into two sets: (1) 70\% is used for training our trust model, and (2) the remaining 30\% is used to compute the accuracy of the constructed model.

\emph{Precision}, \emph{recall}, and \emph{accuracy} \cite{Olson:2008:ADM:1795943} are computed to evaluate the performance of the trust model. Given a set of service provisioning samples and a specific trust level $l$ (e.g., highly trusted), the \emph{precision} for $l$ is computed as the ratio between the number of correctly detected samples as $l$ to the total number of samples detected as $l$ as follows:

\begin{equation}
    Precision_l =\frac{|correct_l|}{|detected_l|}
\end{equation}

\emph{Recall} for $l$ is the ratio between the number of correctly detected samples as $l$ to the actual number of samples under $l$ in the dataset: 

\begin{equation}
    Recall_l =\frac{|correct_l|}{|actual_l|}
\end{equation}

\emph{Accuracy} is the ratio between the number of correctly detected samples as $l$ and correctly detected samples as not $l$ to the total number of samples: 

\begin{equation}
    Accuracy_l =\frac{|correct_l|+|correct\_not_l|}{|samples|}
\end{equation}

Throughout our experiments, we use a 4-layer (input, output, and two hidden layers) fully connected feedforward neural network to train our proposed trust model. The neurons in the input layer are mapped to the answers provided by the MTurk workers. We use five neurons in the output layer, each representing a trust level. More specifically, the neurons in the output layer represent the following trust levels: \emph{highly trusted}, \emph{moderately trusted}, \emph{lowly trusted}, and \emph{not trusted}. We use 32 neurons for each of the hidden layers. The rectified linear unit  (ReLU)  activation function is used in both hidden layers. We use dropout with probability  0.5  on all hidden layers to reduce overfitting.

\subsection{Results}

We train our model to measure the trust level of each service and map it to one of five trust levels discussed earlier. Fig. \ref{fig:pra_original_dataset} and Fig. \ref{fig:pra_interpolated_dataset} show the average precision, recall, and accuracy for the original and interpolated datasets, respectively. For the original dataset, the framework achieves an overall accuracy score of approximately \textasciitilde{80.6\%}. Precision and recall scores are approximately \textasciitilde{80\%} and \textasciitilde{80.4\%}, respectively. The relatively low scores are because of the limited number of samples. Recall that the total number of samples is 5000, of which 70\%, i.e., 3,500, is used to train the model. When the interpolated dataset is used, the framework's accuracy, precision, and recall increased to \textasciitilde{90.5\%}, \textasciitilde{91\%}, and \textasciitilde{89\%}, respectively.

We examine the performance of each of the proposed trust perspectives. The model is retrained using the attributes of one perspective at a time. Each generated model is tested and its accuracy is computed. The results of this experiment is shown in Fig. \ref{fig:per_perspective_accuracy}. The owner perspective achieved the highest accuracy scores (\textasciitilde{70.4\%} and \textasciitilde{75.4\%}, for the original and interpolated datasets) followed by the service perspective (\textasciitilde{55.7\%} and \textasciitilde{64.1\%} for the original and interpolated datasets), while the least accuracy is scored by the device perspective (\textasciitilde{41\%} and \textasciitilde{44\%} for the original and interpolated datasets). It is worth noting that these results can vary depending on the application's type. The current dataset clearly favors the owner perspective to assess trust. Other applications may depend more on the device perspective, e.g., in compute services, properties of the device may have a more importance role than the owner in assessing trust.

The next experiment examines the confidence of our model. Recall that the model is based on a Neural Network. When a Neural Network takes an input, its output layer neurons get activated in varying degrees. The trustworthiness is detected by selecting the trust level that corresponds to the highest activated neuron. Fig. \ref{fig:confidence} shows that the confidence increases until it reaches \textasciitilde{84\%} and \textasciitilde{90\%} for the original and interpolated datasets. It is worth noting that the confidence is increasing at a relatively lower rate compared to the accuracy in Fig. \ref{fig:pra_original_dataset} and Fig. \ref{fig:pra_interpolated_dataset}. The reason is that achieving a high accuracy requires classifying the service into one of the given levels, whereas scoring high confidence requires more epochs to get the weights accurately adjusted.

The efficiency of the framework is examined in the next experiment set. Specifically, we investigate the model training time for different epochs on the two datasets. Fig. \ref{fig:training_time} shows the results for this experiment. As shown in the figure, the training time is linearly proportional to the number of epochs. On the one hand, the model requires \textasciitilde{4} seconds to be trained using 20 epochs and \textasciitilde{21} seconds using 500 epochs for the original dataset. On the other hand, the model training takes \textasciitilde{5} seconds using 10 epochs and \textasciitilde{139} seconds using 500 epochs for the interpolated dataset. 

We investigate the training time in the next set of experiments, Fig. \ref{fig:training_time}. It is clear from Fig. \ref{fig:training_time} that the trust model's training time is linearly proportional to the number of epochs. Using the original dataset, it takes \textasciitilde{4} seconds to train the model using 10 epochs and \textasciitilde{21} seconds using 500 epochs. The interpolated dataset has ten times more samples than the original one. The training time using the interpolated dataset, therefore, takes more time to train; \textasciitilde{5} seconds using 10 epochs and \textasciitilde{139} seconds using 500 epochs.

\begin{figure}
    \centering
    \includegraphics[width=0.40\textwidth]{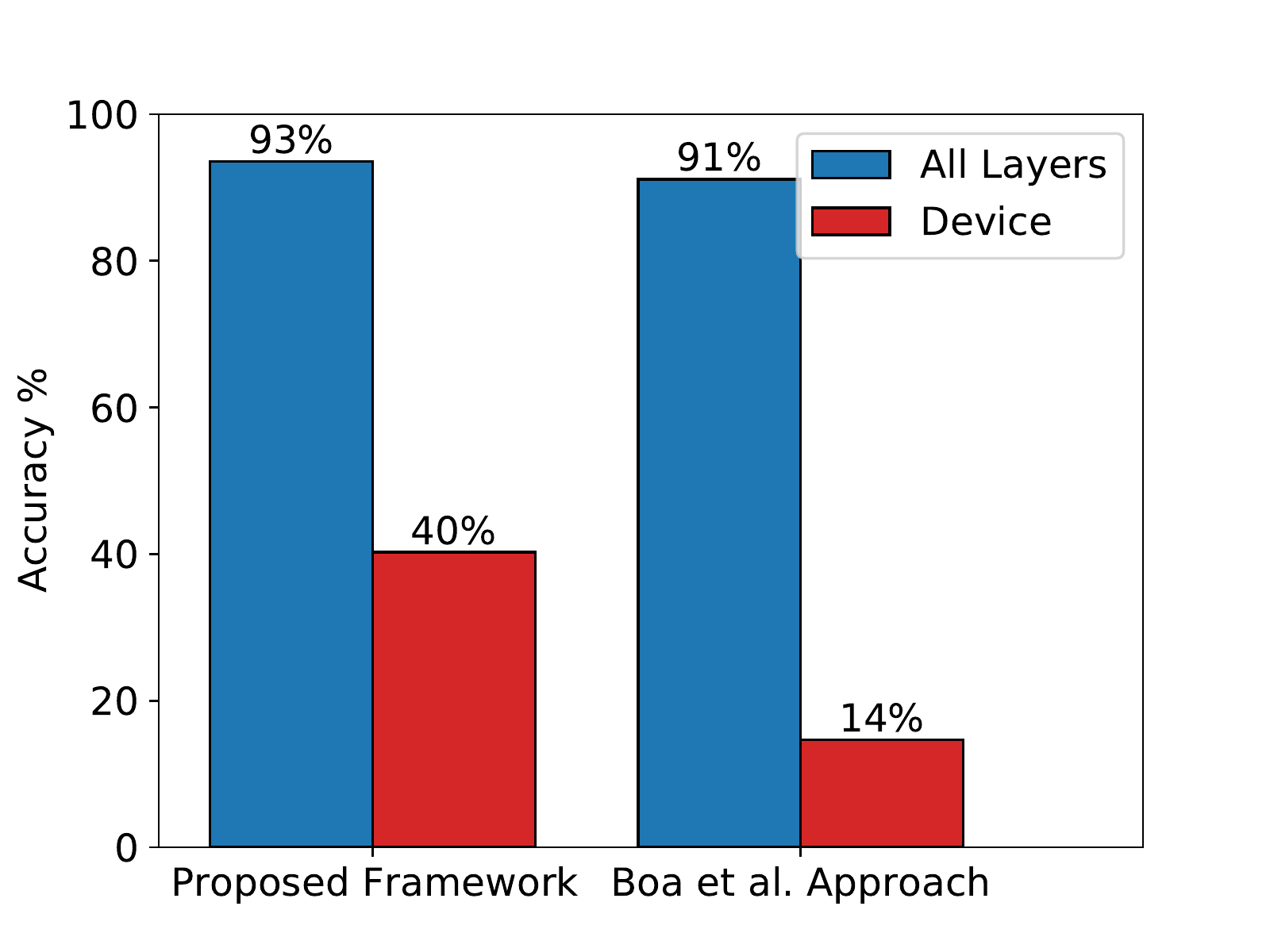}
    \caption{Proposed framework and Bao and Chen's approach accuracies using owner and device properties.}
    \label{fig:comparison}
\end{figure}

The last experiment set compares the accuracy of the proposed framework with the approach in \cite{bao2012dynamic}. As mentioned earlier in Section \ref{section:dataset_description}, we use the SIoT-Yelp dataset for the comparison experiment due to the lack of historical records in our collected MTurk-based dataset. Two experiments have been conducted to compare our approach with Bao and Chen's approach. In the first experiment, we leverage the owner, device, and service perspectives to compute the trust values for our approach. We use the historical records for IoT device owners to compute the trust value for Bao and Chen's approach; as stated in their paper \cite{bao2012dynamic}. The trust levels in the Yelp-SIoT dataset rely heavily on owner attributes. As a result, the significance of leveraging multiple perspectives will not be apparent. Therefore, in the second experiment, we only utilize the \emph{device perspective} to compute the trust value for our approach. Similarly, the \emph{devices'} historical records are used to compute the trust values in Bao and Chen's approach rather than the owners'. Fig. \ref{fig:comparison} shows that both frameworks score high accuracies for the first experiment. The similar result for the two approaches is due to the nature of the dataset. Recall that the ground truth is the extracted ratings from Yelp dataset. Yelp users typically rate businesses based on their quality (reliability) of their services. The approach proposed in \cite{bao2012dynamic} assigns a reputation to services based on their previous ratings (reliability score). Our proposed model also incorporates the reliability when assessing services' trust. It is worth mentioning that while these two approaches give comparable results in this particular scenario, they may not behave similarly when the ground truth considers other aspects of the trust (e.g., the service's IoT device). As a result, our second experiment is conducted, which considers only the device information to compute the trust values. Fig. \ref{fig:comparison} shows the results of the experiment. Both approaches achieve lower accuracy scores compared to the previous experiment; the proposed framework scores 40.24\% while Bao and Chen's approach achieved 14.69\% accuracy. The lower accuracy scores can be attributed to the nature of the dataset; the trust value depends mainly on the owner rather than their devices. Nevertheless, our framework outperforms Bao and Chen's approach by almost three times.

\section{Related Work}
\label{relatedWork}

Assessing the trustworthiness of crowdsourced IoT services is relatively new in the literature. Trust management frameworks generally leverage certain aspects to determine trust, e.g., previous experiences \cite{chen2011trm, Malik2009, saied2013trust, nitti2012subjective, wahab2020endorsement, kantarci2014mobility}, social relations \cite{Cao2015, Caton2014, adali2010measuring, sherchan2013survey}, and  privacy as a means for trust \cite{yu2020privacy}.

A trust model is proposed for evaluating the trustworthiness and reputation of nodes in Wireless Sensor Networks (WSNs) \cite{chen2011trm}. The node’s reputation is computed based on its performance characteristics: packet delivery, forwarding ratio, and energy consumption. The reputation is later used to evaluate the trustworthiness of the node. The trust can be of two types: direct and indirect. Computing the direct trust involves the trustor experience with the trustee node (using the trustee's reputation). Indirect trust is used whenever direct trust cannot be calculated due to the lack of data. The direct trust of the neighbor nodes is used to evaluate the indirect trust. The proposed model uses WSN-specific characteristics, e.g., packet delivery. This prevents the model from being used in IoT applications other than WSNs (i.e., lack of generality).

Another framework is proposed to evaluate the reputation of Web services' providers \cite{Malik2009}. Web services can be consumed by a user or another Web service. The proposed framework allows Web services to share their experiences about other Web services upon consumption. These experiences are then used to form an overall image of the Web services reputations. One key aspect of trust in crowdsourced IoT services is its mutuality. Assuring the trustworthiness of an IoT service provider is not adequate for service consumption. The trustworthiness of the IoT service consumer should be ascertained as well. The framework proposed in \cite{Malik2009} only ensures the reputation of the provider and therefore not suitable for crowdsourcing IoT services.

A centralized trust management system (TMS) is proposed to evaluate the trustworthiness of an IoT device based on its past behavior \cite{saied2013trust}. The TMS has four phases: information gathering, entity selection, transaction and evaluation, and learning. The information gathering phase involves collecting data about the executed services by the IoT device. The data includes the capability of the IoT device, the requester evaluation score, and the time at which the service was obtained. The entity selection phase is where the TMS responds to an IoT device request. The TMS returns the most trustworthy IoT device for the requested service. Performing the task by the IoT service provider happens at the transaction and evaluation phase. The requester gives a score to the provider based on the executed service outcome. In the learning phase, the TMS learns the credibility of the requesters to weight their scores. While TMS performed well according to the conducted experiments, their choice of a centralized solution may bottleneck the flow of the framework. Having a central point where all available IoT devices communicate is impractical due to a large number of IoT devices.

Social IoT network is used to measure the trust between two IoT devices in \cite{nitti2012subjective}. A social IoT network is a type of social networks where nodes are the IoT devices. Relationships between IoT devices indicate one or more of the following relations: similar owner, co-location, co-work, social relation, or brand. Each node computes the trust of its friends. The trust consists of a direct and an indirect trust. The direct trust is computed based on previous experiences and the type of the relation between the two IoT devices. The indirect trust is measured using the friends of an IoT device and their credibility.

A trust management protocol is proposed in \cite{bao2012dynamic, bao2012trust} to assess the trustworthiness of IoT devices. Three trust characteristics are considered: honesty, cooperativeness, and community-interest. Honesty is measured based on the direct observation of an IoT device (high recommendation discrepancy, delays, etc.). Cooperativeness and community-interest are computed using data from social networks. Common friends between two IoT owners indicate high cooperativeness between their IoT devices. Community-interest depends on the number of common communities between two IoT owners.

The work in \cite{rjoub2020bigtrustscheduling} proposes a framework to ensure the trustworthiness of virtual machines (VMs) before task scheduling. Additionally, it proposes a method for categorizing the tasks based on their priority, therefore, high-priority tasks would be assigned with highly trusted virtual machines. The trust of VMs is initially computed by periodically checking if said machines are alive. Moreover, the statistical properties of the machines (CPU, RAM, bandwidth, and disk) is monitored to capture any abnormal activity that might occur. Finally, Markov Chain Monte Carlo Gibbs Sampler (MCMC) is used to combine the aforementioned data and compute the final trust value of a VM.

A framework is proposed in \cite{li2017policy} to detect malicious IoT devices. The framework initially collects data from IoT devices and their contexts. On one hand, context-related data are fed to the "policy management" module of the framework. The role of the policy management module is to set the policies under which IoT devices should report trustworthy data. On the other hand, IoT collected data are forwarded to the "malicious node detection" module, which determines the malice of a device by comparing data from multiple devices and detecting outliers.

The work in \cite{wahab2020endorsement} proposes a framework that bootstraps the trustworthiness of new coming services by collecting endorsements from users. Specifically, the proposed solution has four main phases: endorsement collection, endorsement aggregation, credibility updating, and incentive mechanism. Users provide their recommendations in the endorsement collection phase. The recommendations are aggregated using Dempster-Shafer Theory (DST) to arrive at the final initial trust score. The credibility of the users is computed based on the difference between a user's recommendation and the final aggregated recommendation. During the incentive mechanism, users with higher credibility scores are allowed to offer more endorsements than ones with lower credibility scores.

A framework is proposed for crowdsourcing services to IoT devices based on their mobility and trustworthiness \cite{kantarci2014mobility}. A central authority exists to manage interactions between the service consumer and provider. The trustworthiness of a service provider is computed based on their reputation. Basically, when the central authority receives a task request from the consumer, the task is submitted to multiple service providers. The server then computes the anomalies among the results from the service providers. Service providers with deviated results are marked and their reputation is decreased. The inclusion of a centralized authority may not be practical in IoT environments. Many devices are being deployed continuously. A centralized server can get easily overloaded by the sheer amount of IoT service requests. In addition, the framework does not address how the trustworthiness is evaluated in situations where the service provider is used for the first time, i.e., reputation bootstrapping. 


The work in \cite{azad2020privacy} proposes a decentralized framework that aids in preserving the trustor's privacy in the ``Internet of people'' systems. The framework allows the assignment of weights to the users of the network while ensuring privacy under malicious behaviors. Another framework is proposed in \cite{abououf2019multi} aims at optimizing the execution of tasks in mobile crowdsourcing systems. The framework initially clusters the available tasks based on their geographical locations. Each cluster is then assigned to the optimal group of workers that would maximize the quality of service and minimize the completion time. Another study in \cite{huang2020iot} highlighted the importance of having a trust management system in IoT environments. Specifically, the study collected real data from smarthome IoT devices and analyzed their traffic. It has been found that several devices from known vendors use outdated security measures while others send track users' data for advertising purposes.

It is worth noting that the aforementioned approaches use historical data (i.e., previous experiences) to assess the trust. One key property of IoT environments is their high dynamism in terms of IoT devices deployment. Every day a large number of IoT services are being added. Newly added devices (and, therefore, services) do not have previous records. As a result, previous-experience-based trust management frameworks are not applicable to such environments.


Another set of frameworks in the literature utilize \emph{social networks} to ascertain trust. For example, the framework in \cite{Cao2015}, namely Social WiFi, aims at eliminating the privacy risks accompanied with public WiFi hotspots. An assumption has been made that friends in social networks have a mutual trust between them. Social WiFi utilizes this trust to match hotspot users to trusted hotspot providers. The key part of the work is the integration of social WiFi into the implementation level of the WiFi standard. Specifically, WiFi network discovery and authentication is carried out while considering the social status of the two ends. The proposed framework lacks generality as it can only be used for WiFi hotspot services.

A social compute cloud framework is proposed in \cite{Caton2014} where users in a social network can share and consume services from other users. Some of the main issues in service provisioning are highlighted as follows: trustworthiness, reliability, and availability. The framework tries to overcome these issues by giving users the control over who can use their services and which services they can use (i.e., setting their preferences). The framework leverages the social structure of the social network. The relation types between users (e.g., family, close friends, colleagues, etc) are also utilized to determine the level of trust between them. 

An approach for evaluating the trust between users in social networks is presented in \cite{adali2010measuring}. Behavioral interactions are used to indicate the level of trust (i.e., conversations between users and message propagation). A conversation between two users can indicate a higher level of trust if: (1) it happens many times, (2) it lasts for a long duration, and (3) there is a balanced contribution of messages from both users. The message propagation indicates the willingness of a user $B$ to forward a message received by another user $A$. A large number of forwarded messages reflect a higher trust value for the sender.

Social network-based trust management frameworks may not be sufficient to evaluate the trust between the IoT service provider and consumer. For example, two friends on a social network might not necessitate a mutual trust between them \cite{sherchan2013survey}, and, therefore, cannot only be used as the determining factor for trust. As a result, we investigate to consider other trust-related factors such as the reliability of the IoT service, and the reputation of the IoT service provider and consumer to assess the trust of crowdsourced IoT services.


\emph{Privacy} is also used as a means to ensure trust. For example, the work in \cite{yu2020privacy} assumes an IoT environment where IoT devices (workers/providers) collect data and send them to consumers (requesters). However, such workers might not engage in such an environment due to the inherent loss of privacy in the environment. Therefore, they propose a framework to protect the data, and only authorized consumers can disclose them. More specifically, the framework uses ciphertext-policy attribute-based encryption (CP-ABE), which is a one-to-many encryption method where data can be decrypted by multiple users. While ensuring privacy leads to a trustworthy environment in some applications, it is not the case with IoT crowdsourcing platforms. Other factors (in addition to privacy) can influence the trust of an IoT service. For example, in energy sharing crowdsourcing, the reliability of the service plays a bigger role in determining trust than privacy.

\section{Conclusion}
\label{conculsion}
We proposed an IoT trust management framework that tries to ascertain a level of trust between potential providers and consumers. Specifically, our approach measures the trustworthiness of IoT service providers. The framework is based on a multi-perspective trust model. Each perspective in the model contributes to the overall trust. The perspectives are defined by a set of attributes that are extracted from the inherent characteristics of IoT services. We evaluated our approach in terms of the accuracy of detecting trustful and distrustful IoT services. It is worth noting that our proposed approach evaluates the \emph{service} trustworthiness only (i.e., consumer perspective). Future directions can be investigating the trust level of IoT service consumers. Additionally, our work focused on leveraging trust-related data to evaluate IoT services' trust, while assuming that the privacy and integrity of the data is protected. Our future work focuses on designing a privacy and integrity-preserving framework for storing and exchanging such data between IoT devices.

\section*{Acknowledgment}
This research was partly made possible by DP160103595 and LE180100158 grants from the Australian Research Council. The statements made herein are solely the responsibility of the authors.

\bibliographystyle{IEEEtran}
\bibliography{ref}

\vspace{-1.5cm}
\begin{IEEEbiography}[{\includegraphics[width=1in,height=1.25in,clip,keepaspectratio]{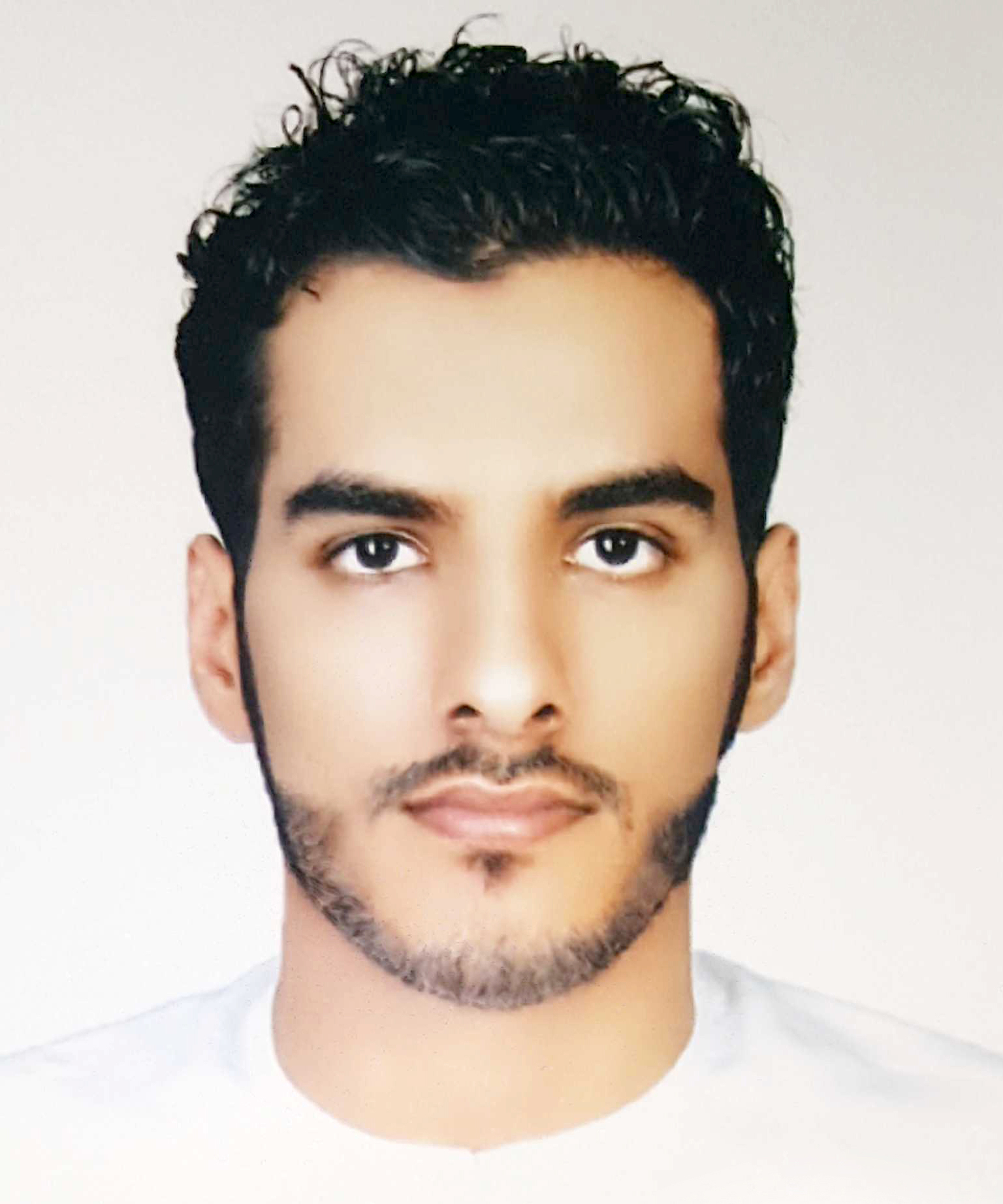}}]{Mohammed Bahutair}
is a PhD student in the School of Computer Science at  the University of Sydney, Australia. He received his bachelor degree in Computer Engineering from Ittihad University, UAE 2012 and his Masters degree in Computer Engineering from University of Sharjah, UAE 2015. His research interests are Machine Learning , Trust,  IoT and Big Data Mining.
\end{IEEEbiography}

\vspace{-1.5cm}
\begin{IEEEbiography}[{\includegraphics[width=1in,height=1.25in,clip,keepaspectratio]{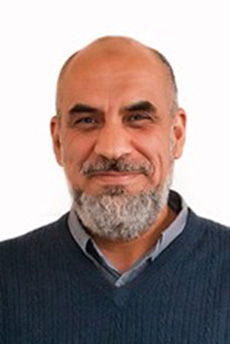}}]{Athman Bouguettaya}
is a Professor at the University of Sydney, Sydney, Australia. He received his PhD in Computer Science from the University of Colorado at Boulder (USA) in 1992. He is or has been on the editorial boards of several journals including, the IEEE Transactions on Services Computing, ACM Transactions on Internet Technology, the International Journal on Next Generation Computing, VLDB Journal, Distributed and Parallel Databases Journal, and the International Journal of Cooperative Information Systems. He has published more than 200 books, book chapters, and articles in journals and conferences in the area of databases and service computing. He is a Fellow of the IEEE and a Distinguished Scientist of the ACM.
\end{IEEEbiography}

\vspace{-1.5cm}
\begin{IEEEbiography}[{\includegraphics[width=1in,height=1.25in,clip,keepaspectratio]{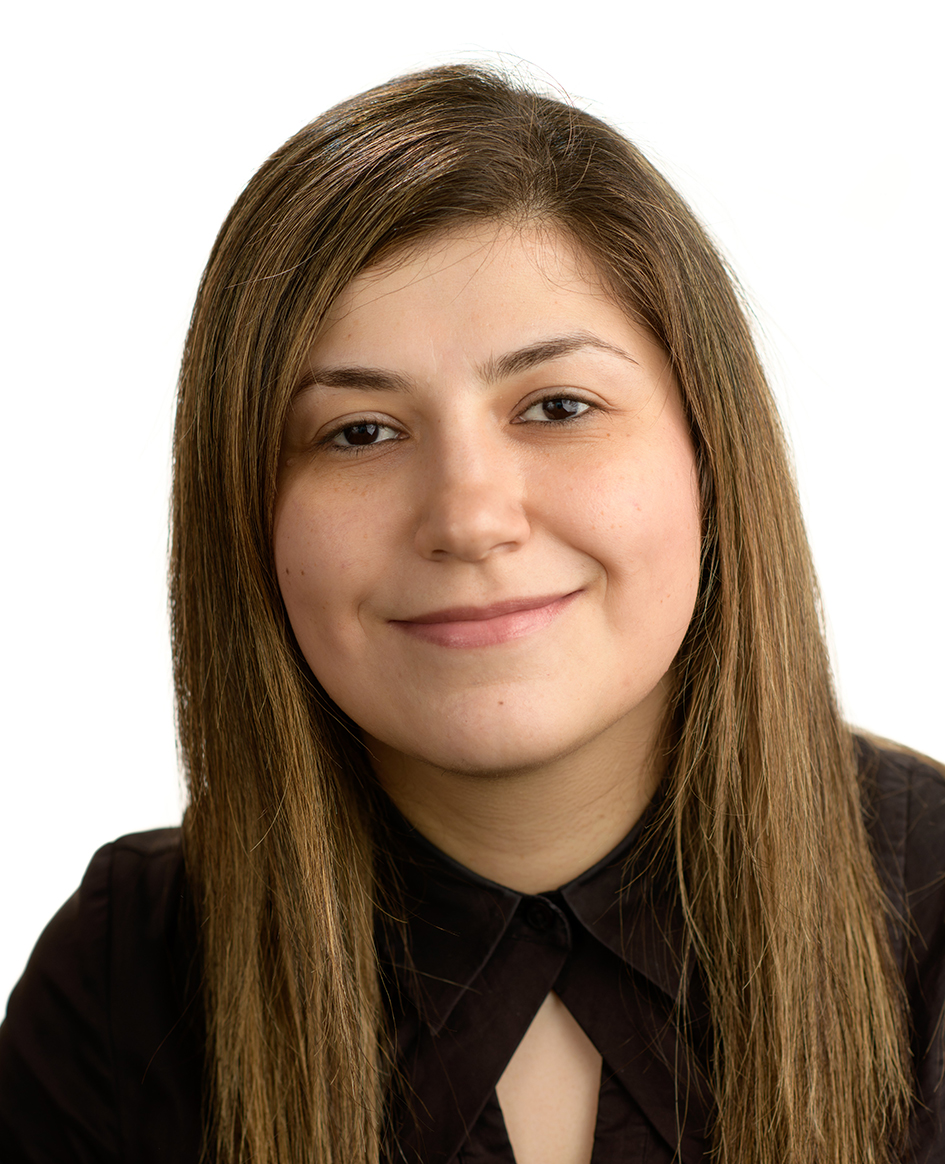}}]{Azadeh Ghari Neiat}
is a lecturer in the School of Information Technology at the Deakin University. She was awarded a PhD in computer science at RMIT University, Australia in 2017. Her current research interests include Internet of Things (IoT), Spatio-Temporal Data Analysis, Mobile Crowdsourcing/Crowdsensing, Big Data Mining, and Machine Learning with applications in the smart city, smart home, and recommender systems.
\end{IEEEbiography}




\end{document}